\documentclass[fleqn,usenatbib]{mnras}

\usepackage[T1]{fontenc}

\DeclareRobustCommand{\VAN}[3]{#2}
\let\VANthebibliography\thebibliography
\def\thebibliography{\DeclareRobustCommand{\VAN}[3]{##3}\VANthebibliography}

\usepackage{graphicx}	
\usepackage{amsmath}	
\usepackage{amssymb}	
\usepackage{natbib}     
\usepackage{cprotect}   
\usepackage{xcolor}      
\usepackage[T1]{fontenc}
\usepackage{indentfirst} 
\usepackage{longtable}  
\usepackage{multirow}   
\usepackage{float}      
\usepackage{fancyvrb}   
\usepackage{tabularx}   
\usepackage[splitrule]{footmisc}
\usepackage{hyperref}   

\usepackage{newtxtext,newtxmath}

\newcommand{\co}{\mathrm{CO}}
\newcommand{\ct}{\mathrm{C}_2}
\newcommand{\ot}{\mathrm{O}_2}
\renewcommand{\cot}{\mathrm{CO}_2}
\newcommand{\cto}{\mathrm{C}_2\mathrm{O}}

\newcommand{\cmfgen}{\textsc{cmfgen}} 
\newcommand{\kms}{\hbox{km$\,$s$^{-1}$}} 
\counterwithout{table}{section}
\counterwithout{equation}{section}

\title[CO in Supernovae]{Carbon Monoxide Cooling in Radiative Transfer Modeling of Supernovae}

\author[C. McLeod et al.]{
Collin McLeod,$^{1}$\thanks{ \url{https://orcid.org/0000-0002-7674-161X} }
D. John Hillier,$^{1}$
Luc Dessart$^{2}$
\\
$^{1}$Department of Physics and Astronomy \& Pittsburgh Particle Physics, Astrophysics, and Cosmology Center (PITT PACC), \\ University of Pittsburgh, 3941 O'Hara Street, Pittsburgh, PA 15260, USA\\
$^{2}$Institut d'Astrophysique de Paris, CNRS-Sorbonne Universit\'e, 98 bis boulevard Arago, F-75014 Paris, France\\
}

\date{Accepted XXX. Received YYY; in original form ZZZ}

\pubyear{2024}

\begin{document}
\label{firstpage}
\pagerange{\pageref{firstpage}--\pageref{lastpage}}
\maketitle

\begin{abstract}

Carbon monoxide (CO) emission has been observed in a number of core-collapse supernovae (SNe) and is known to be an important coolant at late times. We have implemented a chemical reaction network in the radiative-transfer code \cmfgen\ to investigate the formation of CO and its impact on SN ejecta. We calculate two 1D SN models with and without CO: a BSG explosion model at one nebular epoch and a full time sequence (50 to 300 days) for a RSG explosion. In both models, CO forms at nebular times in the dense, inner regions at velocities <2000 \kms\, where line emission from CO can dominate the cooling and reduce the local temperature by as much as a factor of two, weakening emission lines and causing the optical light curve to fade faster. That energy is instead emitted in CO bands, primarily the fundamental band at $\sim 4.5 \mathrm{\mu m}$, which accounts for up to 20$\%$ of the total luminosity at late times. However, the non-monotonic nature of the CO cooling function can cause numerical difficulties and introduce multiple temperature solutions. This issue is compounded by the sensitivity of the CO abundance to a few reaction rates, many of which have large uncertainties or disparate values across literature sources. Our results also suggest that, in many SNe, CO level populations are far from their LTE values. Unfortunately, accurate collisional data, necessary to compute NLTE populations, are limited to a few transitions.



\end{abstract}




\section{Introduction} \label{sec:intro}

Late-time spectra of supernovae (SNe) facilitate the analysis of progenitor and explosion properties by probing the innermost, metal-rich ejecta.  In the nebular phase ($\sim$months after explosion), SN spectra are dominated by emission lines from metals like oxygen, calcium, and iron in the interior of the ejecta, heated by radioactive decay \citep{swartz_late-time_1989,fransson_late_1989,jerkstrand_spectra_2017}.\footnote{In some SNe, the late-time spectra may be powered by shock heating from ongoing interaction with circumstellar medium (CSM).} These lines encode information about the structure and composition of the progenitor, albeit in a complex way. 

Line formation in SNe ejecta involves a diverse set of processes, including energy deposition from $\gamma$-rays, fast-moving non-thermal electrons, and non-Local Thermodynamic Equilibrium (NLTE).  Modeling line formation in SNe thus requires a complex radiative-transfer approach that can take into account many processes simultaneously.  Existing codes for modeling SN spectra include Monte-Carlo codes like \textsc{sumo} \citep{jerkstrand_spectral_2011} and \textsc{jekyll} \citep{ergon_monte-carlo_2018}, as well as full radiative-transfer solvers like \textsc{cmfgen} \citep{hillier_time-dependent_2012}.

SN ejecta have been posited as a potential site for the formation of molecules and dust \citep{gehrz_sources_1989, draine_physics_2011}.  Observations of SN1987A have detected the presence of emission lines from multiple molecules, including carbon monoxide (CO)\footnote{Some authors use ``CO'' when referring to the carbon/oxygen-rich regions or the carbon/oxygen core of the progenitor.  To avoid ambiguity, we use ``CO'' solely to refer to carbon monoxide, and ``C/O'' to refer to carbon and oxygen.} and silicon monoxide (SiO), as well as infrared emission from substantial amounts of dust \citep{spyromilio_carbon_1988, liu_silicon_1994, bouchet_infrared_1993}.  Any dust and molecules present in a stellar atmosphere or nearby environment at the time of explosion would likely be destroyed by the high ($>10000$ K) temperatures and kinetic energies experienced during shock breakout; the presence of molecules at nebular times indicates they must have formed \emph{in situ} as the SN ejecta expanded and cooled, and that they are being actively heated by radioactive decay of $^{56}$Ni and $^{56}$Co \citep{cherchneff_molecules_2011}. Even if molecules are not directly observed, the possibility of core-collapse SNe as significant dust producers \citep{watson_dusty_2015, matsuura_dust_2017, sarangi_dust_2018} requires that molecules like CO and SiO were produced as precursors to dust formation.

Accurate spectral modeling of SNe must include contributions from molecular species in order to recreate observed molecular lines and, more importantly, to take their cooling into account.  Molecules possess rotational and vibrational degrees of freedom which atoms do not, giving them more possible energy states.  These rovibrational energy states have small ($\lesssim 0.5$ eV) energy separations, and transitions between them emit photons in the infrared range.  Transitions between these closely-spaced energy states produce spectral lines that overlap (at velocities $\gtrsim$ tens of \kms ) and are observed as a broad band \citep{herzberg_molecular_1951}. Molecules are very efficient at converting thermal energy into photons, i.e., cooling the gas, as the low-lying energy states are populated by collisional excitation followed by radiative decay.  If molecules form in substantial quantities in SN ejecta, they can alter the ejecta's temperature evolution by accelerating its cooling and affecting the appearance of spectral lines which form in the same region as the molecules (for example, if CO forms in the C/O-rich region, O emission lines will be weakened). 

Nebular spectra of SNe are dominated by forbidden emission lines from the inner, metal-rich ejecta, which originate in the core of the progenitor\footnote{In the absence of interaction with circumstellar medium \citep{dessart_modeling_2022}} \citep{fransson_late_1989,woosley_sn_1988,jerkstrand_spectra_2017,dessart_explosion_2021}.  The emission features in this phase provide direct clues to the progenitor and explosion properties.  For example, the oxygen mass in the ejecta (which can be estimated from the flux of the [\ion{O}{i}] $\lambda \lambda$ 6300, 6364 emission doublet) is correlated with the helium or carbon/oxygen core mass of the progenitor \citep{woosley_evolution_1995,maguire_constraining_2012,dessart_explosion_2021}.  If CO forms in the oxygen-rich region, however, the [\ion{O}{i}] $\lambda \lambda$ 6300, 6364 emission feature will be weakened, lowering estimates of the progenitor core mass derived from it \citep{liljegren_molecular_2022}.  Properly accounting for the cooling effect of CO can mitigate potential systematic errors.

The nature of CO formation in SN ejecta has been investigated using radiative transfer and chemical evolution models covering a range of complexity (though few works have calculated radiation and chemistry simultaneously).  \citet{spyromilio_carbon_1988} estimated a CO mass of $\sim 10^{-4} M_{\sun}$ in SN1987A using a uniform spherical model, assuming the CO level populations followed LTE and that the ejecta was optically thin.  \citet{liu_carbon_1992} extended this treatment to NLTE populations and optically thick CO lines, deducing higher CO masses ($\sim 10^{-3} M_{\sun}$).  

Chemical models have been used to understand the formation of CO in SN ejecta, most with specific application to SN 1987A. \citet{petuchowski_co_1989}, assuming steady-state chemistry and a handful of reactions, concluded that radiative association dominates formation of CO, with destruction dominated by photonic processes. \citet{lepp_molecules_1990} found charge exchange with helium to be among the fastest reactions, but later observations and chemical models \citep{liu_carbon_1992} concluded that very little helium must be present in the oxygen-carbon zone in order to reproduce observed CO abundances ($\sim 10^{-3} M_{\sun}$). While these models all assumed steady-state chemistry, later work \citep{cherchneff_molecules_2011} suggested that time-dependent chemistry is necessary. Chemical mixing (or the lack thereof) also plays an important role in molecular formation, as the presence of $^{56}$Ni in the oxygen-carbon zone can quickly dissociate molecules (directly via gamma rays or indirectly via heating the gas), limiting molecular formation to only regions with a very low abundance of radioactive isotopes \citep{ono_impact_2023}.

Recently, \citet{liljegren_carbon_2020} included CO formation (assuming steady-state chemical equilibrium) and emission in the Monte-Carlo spectral synthesis code \textsc{sumo}, investigating its effects in Type II (SN1987A) and Type Ibc SNe \citep{liljegren_molecular_2022}, finding that molecules form in significant quantities ($\geq 10^{-3} M_{\sun}$) in the ejecta of both SN types and accelerate the cooling.

We undertake a similar investigation by including CO in the radiative-transfer code \textsc{cmfgen}. While \textsc{sumo}, \textsc{jekyll}, and many other codes compute the radiative transfer using a Monte-Carlo approach, \cmfgen\ solves for the moments of the radiation field numerically. \cmfgen\ also has the advantage of broad applicability, having been used to model Type Ia \citep{dessart_critical_2014,blondin_one-dimensional_2013}, Ib/c \citep{dessart_supernovae_2020}, and Type II SNe \citep{dessart_quantitative_2005,jacobson-galan_final_2022} as well as stars, including O \citep{martins_new_2005,bestenlehner_vlt-flames_2014}, B \citep{thompson_iron_2008,keszthelyi_effects_2021}, and Wolf-Rayet \citep{de_marco__2000,massey_modern_2014} stars. While early work using Monte-Carlo codes required simplifying assumptions about the ionization structure or excited state populations \citep{mazzali_application_1993}, more recent work (e.g., using \textsc{jekyll}) treats NLTE fully and includes a very similar set of physical processes to \cmfgen . It is useful to compare the results of different approaches applied to the same complex problem.

With the launch of the James Webb Space Telescope, high-quality infrared observations of core-collapse SNe should become available in the coming years. While many chemical models have been produced to explain the formation of CO in SN ejecta, very few works have explicitly connected chemical evolution and spectral observations by including molecules in radiative transfer calculations. Moreover, CO has been detected in a variety of SN types, including Type IIP \citep{kotak_early-time_2005, kotak_dust_2009,rho_near-infrared_2018}, IIL \citep{yuan_450_2016}, Ic \citep{banerjee_near-infrared_2016, stritzinger_carbon-rich_2023,rho_near-infrared_2021,drout_double-peaked_2016}, and IIn \citep{gerardy_detection_2000,fassia_optical_2001}, suggestive that formation of molecules may be commonplace in SNe. Thus it is crucial to compare chemical reaction models with observations of a wide variety of SN types. 

Using two simple SN models, we investigate the \emph{in situ} formation of CO in SN ejecta via a self-consistent set of chemical reactions, and consider the impact of ejecta cooling due to rovibrational emission from CO. The SN models used in this work feature extensive microscopic mixing of composition and are intended as a first demonstration of the impacts of molecules in \textsc{cmfgen}.  More physically realistic models (featuring only macroscopic mixing) require longer computation times \citep{dessart_radiative-transfer_2020}, but will be presented in future work. Section \ref{sec:cmfgen} describes the function of \textsc{cmfgen} and the alterations made to it which allow for CO formation and emission, \S\ref{sec:models} outlines the details of the SN models used in our investigation, and \S\ref{sec:results} discusses our results.  Section \ref{sec:cooling} details the behavior of the CO cooling function which warrants particular attention, \S\ref{sec:context} compares our results to other chemical models of SNe and to observations of CO in Type II SNe, and \S\ref{sec:conclusion} contains our conclusions.

\cprotect\section{\textsc{cmfgen}} \label{sec:cmfgen}
%
\textsc{cmfgen} is a non-LTE radiative transfer code originally designed to model hot stars with winds \citep{hillier_cmfgen_2001,hillier_iterative_1990}. It uses a Newton-Raphson technique to solve for the moments of the radiative transfer equation simultaneously with the rate equations and radiative equilibrium equation in the comoving frame.  \textsc{cmfgen} has been adapted to simulate SNe, treating time-dependent effects \citep{hillier_time-dependent_2012}, deposition of radioactive decay energy, and non-thermal processes \citep{li_non-thermal_2012}. 

\textsc{cmfgen} is a one-dimensional code, which necessitates the use of some simplifying assumptions. We assume that the SN ejecta is spherically symmetric and expanding homologously. \textsc{cmfgen} includes multiple approaches for chemical mixing: standard boxcar smoothing \citep{dessart_supernova_2010} or a shuffled-shell approach \citep{dessart_radiative-transfer_2020} which allows for macroscopic mixing but not microscopic mixing. \footnote{Fluid instabilities during the explosion produce large bubbles or ``fingers" of metal-rich material from the core (which later expand as a result of heating from radioactive $^{56}$Ni and $^{56}$Co). These $^{56}$Ni-rich regions break spherical symmetry and extend out to large velocities but may not introduce significant microscopic mixing between layers \citep{fryxell_instabilities_1991,kifonidis_non-spherical_2003,wongwathanarat_three-dimensional_2015}. Observed line profiles in many objects suggest minimal mixing between ejecta shells: for example, microscopic mixing of calcium into the O-rich shell would produce weaker oxygen lines and stronger calcium lines than are typically observed \citep{dessart_radiative-transfer_2020,dessart_explosion_2021}. Direct imaging of SN1987A also indicates little mixing between C/O- and Si/O-rich shells \citep{abellan_very_2017}. For further discussions of mixing (which is not the focus of this work), see \citet{dessart_radiative-transfer_2020-1}, \citet{ono_impact_2023}, and \citet{dessart_radiative-transfer_2020}.} In this work, we use the simpler boxcar approach to reduce complexity.

Each ion included in a model adds a number of rate equations to the total system.  Every included energy level is associated with one equation of the form

\begin{equation}\label{eqn:rate_eqn1}
 \rho \frac{D n_i / \rho}{Dt} = \sum_j (n_j R_{ji} - n_i R_{ij})
\end{equation}

\noindent where $\rho$ is the mass density, $n_i$ the number density of the state $i$, and $R_{ij}$ the rate of all processes from state $i$ to state $j$. $R_{ij}$ includes photoionization, recombination, line absorption/emission, etc. \citep[see][]{hillier_time-dependent_2012}. In general, these processes couple energy states within a particular ion and neighboring ionization states. When molecular reactions are included (see \S \ref{sec:chem_rxns}), $j$ can refer additionally to states in different species (e.g., the equation for the \ion{O}{I} ground state will include terms for rates involving CO energy levels). 

To reduce the total number of equations which must be solved, \textsc{cmfgen} makes use of \emph{superlevels}, in which multiple energy levels are grouped together. Such a group of (usually closely-spaced) energy levels constitutes one superlevel, which is associated with a single equation. The relative populations of levels grouped within a superlevel are assumed to be in LTE. For a more complete discussion of superlevels, see \citet{hillier_treatment_1998}.  Each atomic species (i.e., hydrogen, helium, carbon, etc.) also gives rise to a constraint equation of the form of equation \ref{eqn:constraint_1} which specifies that the total amount of a particular species remains fixed:

  \begin{equation}
  \sum_{j=1}^N \frac{D N_j}{Dt} = 0
  \end{equation}\label{eqn:constraint_1}

\noindent for $N_j$ the population of ionization state $j$, referred to as the number conservation equation. This constraint must be modified when molecules are included; see section \ref{sec:other_changes}.

\subsection{Carbon Monoxide} \label{sec:cmf_co}

We add carbon monoxide to \textsc{cmfgen}, and treat it similarly to an atom or ion. Unlike atoms, CO can form \emph{in situ} in the SN ejecta via chemical reactions.

\subsubsection{Chemical Reactions}\label{sec:chem_rxns}

We allow for the formation of CO in the ejecta by way of a set of chemical reactions, with data obtained from the \verb|UMIST| database \citep{mcelroy_umist_2013}.  A full list of reactions is included in table \ref{table:rxns}. The change to a species from each chemical reaction is included in the relevant kinetic equation. These equations are then iteratively solved with a Newton-Raphson approach in the full matrix solution.  In a full time sequence of SN models, we calculate the time-dependent chemical equilibrium (using equation \ref{eqn:rate_eqn1}) to determine the CO population. However, we also compute some time-independent models in which we assume steady-state chemical equilibrium (modifying equation \ref{eqn:rate_eqn1} to eliminate the time-dependent term). The rate of each reaction is calculated using an equation of the form:

\begin{equation}
r = \alpha \; n_{R_1} n_{R_2} \left( \frac{T}{300K} \right)^{\beta} \exp \left( -\gamma / T \right)
\label{eqn:rate}\end{equation}

\noindent where $n_{R1}$ and $n_{R2}$ are the number densities of the reactants, $T$ is the local electron temperature, and $\alpha$, $\beta$, and $\gamma$ are constants specific to each reaction.  When solving equation \ref{eqn:rate_eqn1}, the rate $r$ replaces either $n_i R_{ij}$ or $n_j R_{ji}$, depending on whether the state $n_i$ is a reactant or product. Reverse reaction rates are unavailable for many of the reactions we consider, and we do not attempt to compute reverse reaction rates when they are not available.  Data on the dependence of reaction rates on the electronic/rovibrational state are likewise missing.  We make the following assumptions regarding reactant/product energy states and reaction rates:

\begin{enumerate}
\item Only \textit{ground term} states in atomic species participate in chemical reactions (as reactants or products). Generally this is the ground LS term, which can consist of multiple energy states. For example, for neutral oxygen (\ion{O}{i}), this is the $2\rm{s}^22\rm{p}^4$ $^3\rm{P}$ term, which has 3 J states. Other than determining participation in chemical reactions, details of the individual states (energy, angular momentum, etc.) have \textit{no impact} on the rate of the reaction, which depends only on the population of the state. Higher energy states are ignored when calculating molecular reaction rates.
\item All molecular energy states can participate in chemical reactions, and the rate has \textit{no dependence} on energy states of molecular reactants/products.
\item Potential atomic product states are populated proportional to their statistical weights. For example, the reaction $\mathrm{C}_2 \; + \; \mathrm{O}^+ \; \rightarrow \; \mathrm{C}_2^+ \; + \; \mathrm{O}$ populates all 3 J states of the OI ground term:  it populates the $J=2$ state ($g=5$) at $\frac{5}{9}$ of the total reaction rate, the $J=1$ state ($g=3$) at $\frac{3}{9}$ the total rate, and the $J=0$ state ($g=1$) at $\frac{1}{9}$ the total rate.
\end{enumerate}

We neglect potential cooling due to these chemical reactions, as the rates are very small compared to other heating/cooling processes.\footnote{In the CO-dominated region of our BSG model, radiative association dominates the formation of CO.  The cooling rate (enthalpy change) from this reaction is $\sim 5 \times 10^{-12} \; \rm{erg} \; \rm{s}^{-1} \; \rm{cm}^{-3}$, compared to a CO line cooling rate of $\sim 1.6 \times 10^{-7} \; \rm{erg} \; \rm{s}^{-1} \; \rm{cm}^{-3}$ and a net cooling rate of $\sim 3.7 \times 10^{-9} \; \rm{erg} \; \rm{s}^{-1} \; \rm{cm}^{-3}$.  This reaction exceeds by at least a factor of 100 the cooling rate from any other molecular reaction.}

Creating a complete reaction network necessitates the inclusion of additional molecular species, as CO may be formed via multi-step processes that include reactions such as $\mathrm{C}_2 \; + \; \mathrm{O} \; \rightarrow \; \mathrm{C} \; + \; \mathrm{CO}$.  The species $\mathrm{C}_2$, $\mathrm{O}_2$, $\mathrm{CO}_2$, and $\mathrm{C}_2\mathrm{O}$ are included in \textsc{cmfgen}. They are, however, excluded from radiative transfer calculations. All molecules other than CO are treated using only a single energy state, no transitions, and no photoionization pathways. We currently limit the reaction network to carbon and oxygen--keeping the reaction network as simple as possible and its computational cost low while we develop our method. Future work will include a much larger network with additional molecules\footnote{Our results from this study indicate that the computation time for chemical reactions is insignificant compared to the radiative transfer calculation, so a larger reaction network should not significantly impact the total program runtime in future work.}.

Molecules can also be destroyed by collisions with high-energy electrons.  In SNe, some electrons are accelerated via inverse-Compton scattering with gamma rays produced by radioactive decay of (primarily) $^{56}$Ni and $^{56}$Co.  These fast-moving electrons are often referred to as Compton electrons, and can collide with molecules and dissociate them into constituent atoms/ions--we include three pathways (Table \ref{tab:compton_rxns}).

\begin{table}
  \begin{center}
    \caption{Compton (non-thermal) reaction processes included in \textsc{cmfgen}. Data for cross-sections come from \citet{itikawa_cross_2015}, by way of the LXCAT database \citep{pancheshnyi_lxcat_2012}. Data were extrapolated using a power law to match the sampling of the non-thermal electron spectrum in \textsc{cmfgen}.}
    \begin{tabular}{||c|c|c|c||}
      \hline
      ID & \multicolumn{3}{|c|}{Reaction} \\
      \hline
      1 & $\mathrm{CO} + e^+_{non-therm}$ & $\rightarrow$ & $\mathrm{C} + \mathrm{O} + e^-$ \\
      2 & $\mathrm{CO} + e^+_{non-therm}$ & $\rightarrow$ & $\mathrm{C}^+ + \mathrm{O} + e^- + e^-$ \\
      3 & $\co + e^+_{non-therm}$ & $\rightarrow$ & $\mathrm{C} + \mathrm{O}^+ + e^- + e^-$ \\[1ex]
      \hline
    \end{tabular} \label{tab:compton_rxns}
  \end{center}
\end{table} 

\noindent We calculate the rate of destruction of CO according to

\begin{equation}
r = n_{\co} \int_0^{\infty} \sigma (v) \; v \; \frac{d n_e}{dv} dv
\end{equation}\label{eqn:comp_rate}

\noindent where $n_{\co}$ is the total number density of CO, $\sigma (v)$ the energy-dependent cross-section for interaction, $v$ the electron velocity, and $\frac{d n_e}{dv}$ the non-thermal electron spectrum (distribution of non-thermal electrons as a function of velocity). \textsc{cmfgen} explicitly calculates the spectrum of non-thermal electrons present in the ejecta using the Spencer-Fano equation \citep{li_non-thermal_2012}. The cross-sections for interaction were taken from \citet{itikawa_cross_2015}, fitted with a power law, and interpolated to the same grid as the electron spectrum.  These reactions have no dependence on the temperature nor on the number density of thermal electrons.

\subsubsection{CO Levels and Lines}

We obtain rovibrational level and line lists for CO ($^{12}\mathrm{C}^{16}\mathrm{O}$ isotopologue) from \citet{li_rovibrational_2015}. We include vibrational levels up to $v=10$ and rotational levels up to $J=75$.  These limits were selected based on available collisional data for CO--data from \citet{ogloblina_electron_2020} covers $v=0$ up to $v=10$. The resulting data consist of 836 rovibrational levels and 9075 transitions, including level energies, quantum numbers, and statistical weights, along with transition wavelengths, oscillator strengths, and Einstein A values.  We include only the ground electronic state for CO ($X^1 \Sigma^+$), assuming that excited electronic state populations are negligible under the conditions in which CO forms. In LTE at $\approx$5000K, the first excited electronic state of CO, $a^3\Pi$, which lies $\sim 6$ eV above the ground state, would have a Boltzmann factor of $\sim 10^{-7}$, i.e., a negligible population \citep{essenhigh_energy_2005}.  These data are included in \textsc{cmfgen} in the same format as atomic energy levels. When calculating the radiative energy loss due to CO emission, we account for all radiative transfer and optical depth effects. See \S\ref{sec:cooling} for a discussion of how we characterize the cooling effect of CO line emission.

In \textsc{cmfgen}, CO is treated as any other atom, except without a fixed abundance (see \S \ref{sec:other_changes}).  It is treated using the full line-blanketing approach. We make no changes to the radiative transfer in \textsc{cmfgen} aside from the inclusion of CO lines.

We include two approaches to the CO level populations. The total abundance of CO is determined by the chemical reactions as described in \S\ref{sec:chem_rxns}, and is independent of the level population approach.  In approach 1, we assume that all CO level populations are in LTE, and they are treated in the equations using a single superlevel \citep[see][for discussion of superlevels]{hillier_treatment_1998}. In approach 2, we allow CO level populations to deviate from LTE, using electron-collision cross-sections from \citet{ogloblina_electron_2020} to compute the populations.  The cross-sections from \citet{ogloblina_electron_2020}, which cover vibrational transitions of the form $v|0 \rightarrow v^{\prime}$, were integrated with a Maxwell-Boltzmann distribution to obtain thermal collision rates for vibrational transitions, then scaled using the relation given in \citet{chandra_collisional_2001}:

\begin{equation}
\frac{C(T|v \rightarrow v_1)}{C(T|v \rightarrow v_2)} = \frac{2v_1+1}{2v_2+1}
\end{equation}\label{eqn:chandra_relation}

\noindent to cover all possible vibrational transitions.  These collision rates, together with radiative de-excitation rates (quantified with Einstein coefficients) determine the relative vibrational-state populations.  The level populations are calculated using a single superlevel for each vibrational state, assuming that all rotational states within a given vibrational state have the same departure coefficient.  The validity of the full LTE assumption is discussed in \S\ref{sec:results} and \S\ref{sec:context}.

\subsubsection{Other changes to \textsc{cmfgen}} \label{sec:other_changes}

In the SN ejecta, the total number density of a given element is fixed, i.e., all oxygen atoms remain oxygen atoms over time (though some may be neutral \ion{O}{i}, some ionized \ion{O}{ii}, etc.).\footnote{Via radioactive decay, some elements can change: $^{56}$Ni decays to $^{56}$Co over time. This process, however, has a fixed half-life and does not depend on the local conditions in the ejecta, while the population of CO does.}  This is quantified in the number conservation equation.  The quantity of molecular species, however, is not fixed --- CO does form over time, and can be destroyed. We modify the oxygen and carbon conservation equations to account for molecules.  The new conservation equation specifies that the total quantity of carbon (oxygen) is fixed, with some in the form of atomic carbon (oxygen) ions, and some in the form of molecules:

\begin{equation}
n_{carbon} = \sum_{i=0}^{i_{max}} n_{C^{i+}} + n_{\co} + 2n_{\ct} + n_{\cot} + 2n_{\cto}
\end{equation}\label{eqn:new_number_cons}

\section{Model Setup}\label{sec:models}

We calculate two different SN models: a red supergiant (RSG) model \citep{hillier_photometric_2019} and a blue supergiant (BSG) model \citep{dessart_supernovae_2019}, each with multiple approaches.  Both models are based on a progenitor with a zero-age main sequence (ZAMS) mass of 15 $M_{\sun}$. First each model is calculated without allowing for the formation of molecules (NOCO).  Then each model is recalculated allowing for the formation of CO with level populations and line emission treated in LTE (LTECO).  The BSG model is also recalculated allowing CO vibrational state populations to deviate from LTE (NLTECO). Both models are based on simulations of 15 $M_{\sun}$ progenitors, evolved with the stellar evolution code \textsc{mesa} \citep{paxton_modules_2018} and then exploded using the radiation hydrodynamics code \textsc{v1d} \citep{livne_implicit_1993}. Both models feature a mixed composition, which is performed using a boxcar method following the initial explosion simulation \citep[following the method of][]{dessart_supernova_2010}. Mixing allows CO to form throughout the ejecta, wherever carbon and oxygen are present. 

\begin{enumerate}
\item BSG:  This model is based on a 15 $M_{\sun}$ BSG progenitor which produces a Type II-peculiar SN (observational examples of this type include SN1987A, SN2000cb, and SN2009E) \citep{arnett_supernova_1989, kleiser_peculiar_2011, pastorello_sn_2012}.  Full details of the model can be found in \citet{dessart_supernovae_2019}, from which we use the ``a4'' model.  This model is evolved at sub-solar metallicity and scaled to match the $^{56}$Ni mass (0.08 $M_{\sun}$) observed in SN1987A \citep[see][]{arnett_supernova_1989, spyromilio_carbon_1988}.  After the initial explosion, all species are macroscopically mixed throughout the ejecta \citep[see][]{dessart_supernovae_2019}, creating a strongly mixed composition.  At 1 day after explosion, the \textsc{v1d} model is remapped to \cmfgen\ assuming homologous expansion (which is a good approximation at that time).  This model is evolved to 330 days without the inclusion of molecules.  At that epoch, we remove all time-dependent terms (which are small) and add molecules, assuming the ejecta is now in chemical equilibrium. In our molecular analysis, we consider only this single snapshot of the evolution.  Fig. \ref{fig:a4_comp} shows the composition profiles in this model.

  \begin{figure}
    \begin{center}
      \makebox[\linewidth][c]{\includegraphics[width=1.1\linewidth]{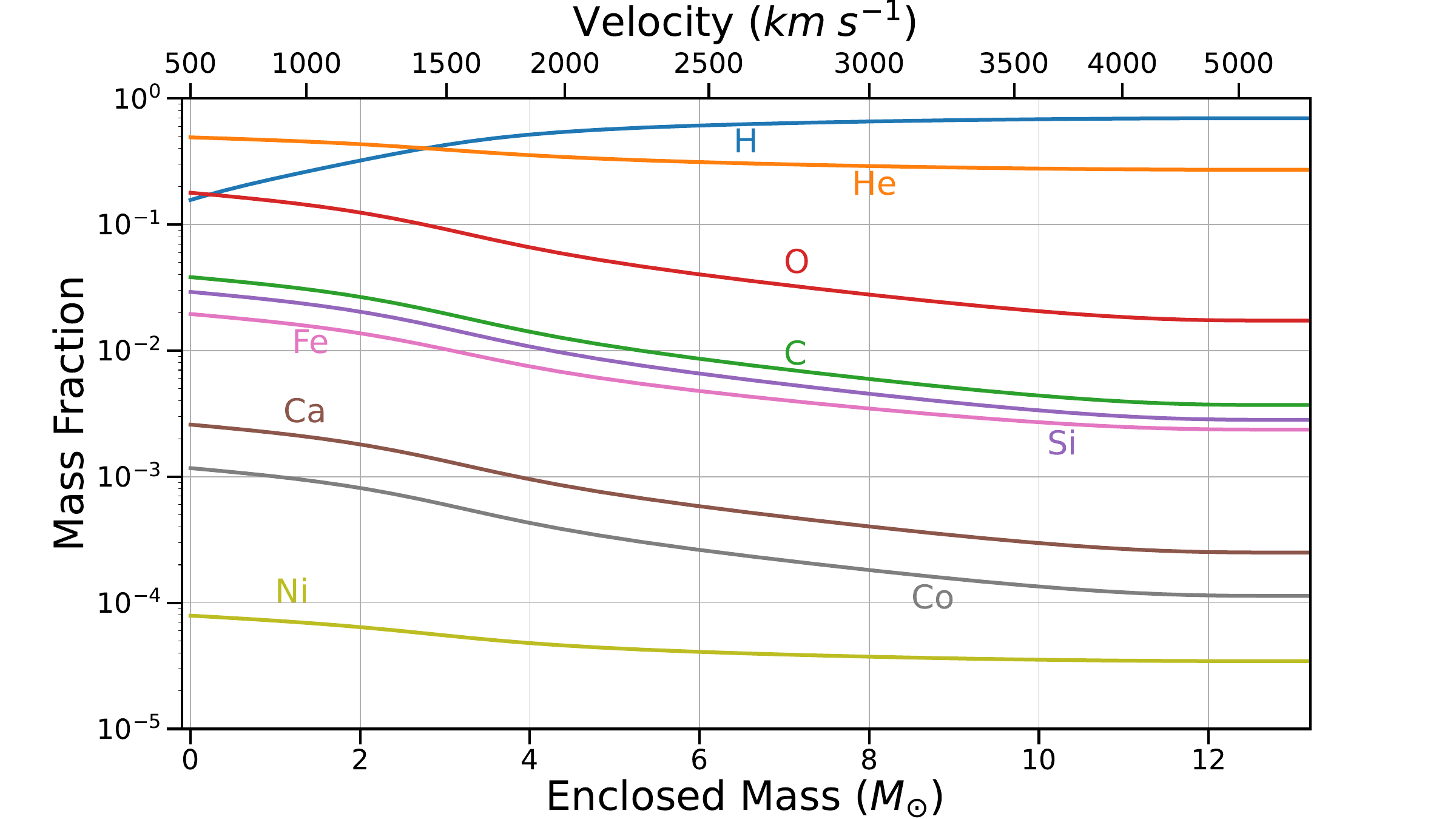}}
    \caption{Mass fractions of important elements at 330 days post-explosion in the BSG model, which is microscopically mixed using a boxcar. This model has a total ejecta mass of 13.22 $M_{\sun}$, of which 7.31 $M_{\sun}$ is hydrogen and 0.0843 $M_{\sun}$ $^{56}$Ni.  The interior of this model is dominated by helium and oxygen, with a smaller proportion of carbon.}
    \label{fig:a4_comp}
    \end{center}
  \end{figure}
  
\item RSG: This model is based on a 15$M_{\sun}$ RSG progenitor which explodes to produce a typical Type IIP SN (examples include SN1999em, SN2004et, and SN2017eaw \citep{leonard_distance_2002,kotak_dust_2009,rho_near-infrared_2018}).  Specific details of the model can be found in \citet{hillier_photometric_2019}, from which we use the ``x2p0'' model.  This model is calculated from a progenitor which retains a significant ($>9M_{\sun}$) hydrogen envelope at the time of explosion.  The explosion model produces 11.87 $M_{\sun}$ of ejecta, of which 0.036 $M_{\sun}$ is $^{56}$Ni, which provides the heating at late times. After explosion, the composition is mixed using a boxcar algorithm, with weaker mixing than in the BSG model.  We do not include CO until 50 days after the explosion. At early times, the high temperatures in the oxygen-rich region should result in very little molecular formation. 50 days was chosen so that CO begins to form before $\sim 100$ days, per the earliest detections of CO in SN1987A \citep{spyromilio_carbon_1988}. The model is evolved until 300 days post-explosion, including all time-dependent terms and the effect of CO cooling on its evolution. Fig. \ref{fig:x2_comp} shows composition profiles for the RSG explosion model.

 \begin{figure}
    \centering
    \makebox[\linewidth][c]{\includegraphics[width=1.1\linewidth]{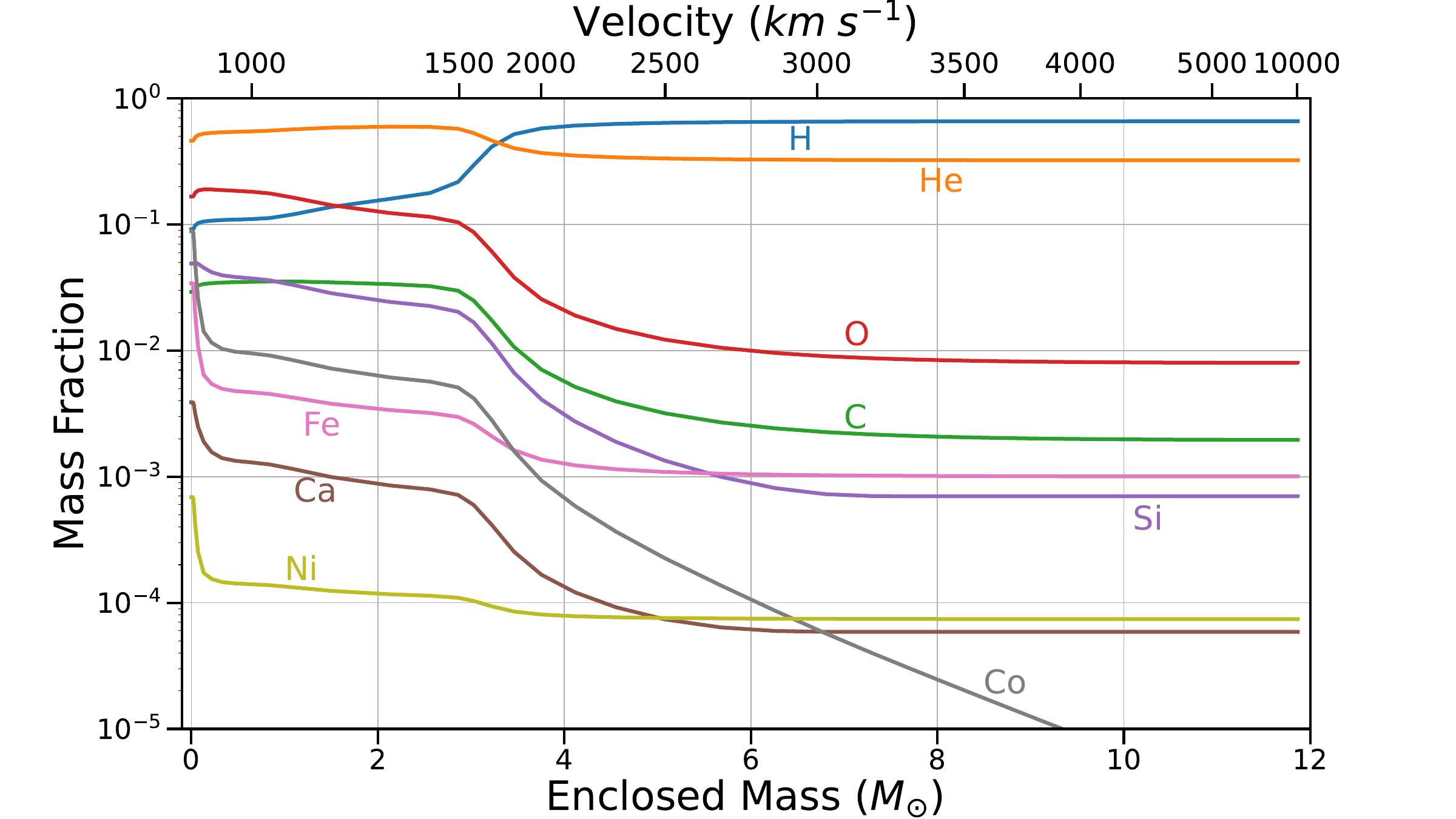}}
    \caption{Mass fractions of important elements at 46 days post-explosion in the RSG model, which is microscopically mixed using a boxcar. The innermost region is dominated by helium, oxygen, and silicon, followed by a shell of primarily helium, oxygen, and carbon.  The outermost $\sim 9 M_{\sun}$ is dominated by hydrogen.}
    \label{fig:x2_comp}
  \end{figure}

\end{enumerate}

\begin{figure}
 \centering
 \makebox[\linewidth][c]{\includegraphics[width=1.1\linewidth]{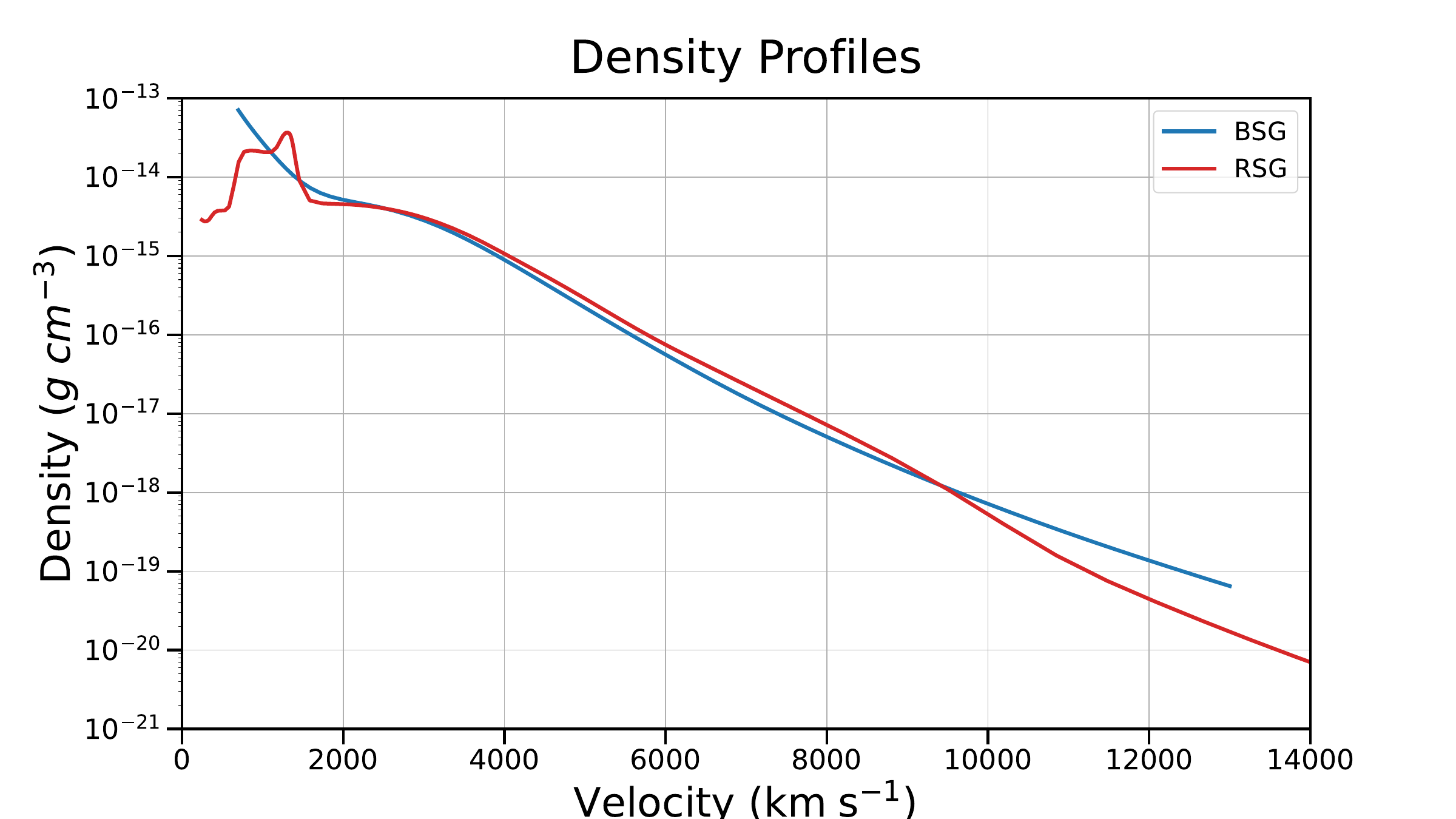}}
 \caption{Density profiles of the BSG and RSG models at the same epoch (300 days). Both models have a power-law density distribution at high velocities, but they differ at low velocities.  Near the inner parts of the ejecta, the RSG model is more stratified, with a dense C/O shell. The primary difference between the models is that the RSG progenitor has a much larger radius at the time of the explosion.}
 \label{fig:rho_profs}
\end{figure}

Both models are computed from a 15 $M_{\sun}$ progenitor but with different explosion properties, notably the density structure and progenitor radius at the time of the explosion (Fig. \ref{fig:rho_profs}).

\section{Results and Discussion} \label{sec:results}

\subsection{CO Formation}

In our RSG model, $1.76 \times 10^{-4} M_{\sun}$ of CO has formed by 330 days, while our BSG model (at 330 days) contains $6.1 \times 10^{-5} M_{\sun}$ of CO. In both models, CO forms preferentially in conditions of high density and low temperature.  It is formed in the inner regions of the ejecta (<2000 \kms\ in our BSG model, <5000 \kms\ in our RSG model), with the majority of CO forming after $\sim$100 days post-explosion in our RSG model. The formation of CO depends strongly on temperature:  in regions where the temperature exceeds $4000-4500$ K (depending on the local density), negligible quantities of CO form, as it is quickly destroyed by non-thermal electrons and charge exchange with oxygen.  In lower-temperature regions, CO can form in significant quantities, though the local number density of CO never exceeds $\sim$1 per cent of the carbon number density in any region within our models. A temperature below $4000$K is not sufficient for CO to form in large quantities; relatively high densities are also required (though we do not explore enough of the parameter space to make a quantitative estimate of a density threshold).  We do not observe any scenario in which most of the carbon or oxygen becomes locked up in molecules, even at late ($\geq 300$ days post-explosion) times and low ($\sim 1000$K) temperatures. The equilibrium quantity of CO increases as density increases or temperature decreases.


The equilibrium population of CO is dictated by a few reactions.  The radiative association reaction $\mathrm{C} \; + \mathrm{O} \; \rightarrow \; \co \; + \; \gamma$ typically dominates the formation of CO, though the charge exchange reaction $\mathrm{CO}^+ \; + \; \mathrm{O} \; \rightarrow \; \co \; + \; \mathrm{O}^+$ can become important when the density of neutral oxygen is high. Charge exchange with oxygen: $\co \; + \; \mathrm{O}^+ \; \rightarrow \; \mathrm{CO}^+ \; + \; \mathrm{O}$ typically dominates the destruction of CO, but dissociation by non-thermal electrons ($\co \; + \; e^{-}_{non-therm} \; \rightarrow \; \mathrm{C}/\mathrm{C}^+ \; + \; \mathrm{O}/\mathrm{O}^+$) can also contribute significantly, especially at lower densities.  CO ionization balance is dictated by charge exchange with O and $\mathrm{O}^+$, as the reaction rates exceed by orders of magnitude the CO photoionization/recombination rate. 


Our prescription for chemical mixing included in both models likely impacts where and how much CO forms. The chosen boxcar approach allows CO to form throughout the ejecta, following the distribution of carbon and oxygen. The shuffled-shell approach, while more realistic, requires higher spatial resolution and thus longer computation times. Our initial results using boxcar mixing likely overestimate the spatial extent of CO formation, but possibly underestimate the local density of CO that could form in dense, carbon/oxygen-dominated shells. Boxcar mixing also means that CO is distributed throughout the ejecta with all other elements, so its cooling effect impacts not only carbon and oxygen but also hydrogen and other species. With a stratified shell model, we would expect CO to affect only the species in the carbon/oxygen-rich shell, leaving the outer, hydrogen-rich layers unaffected.

\subsubsection{Formation over time}

\begin{figure}
  \centering
  \makebox[\linewidth][c]{\includegraphics[width=1.1\linewidth]{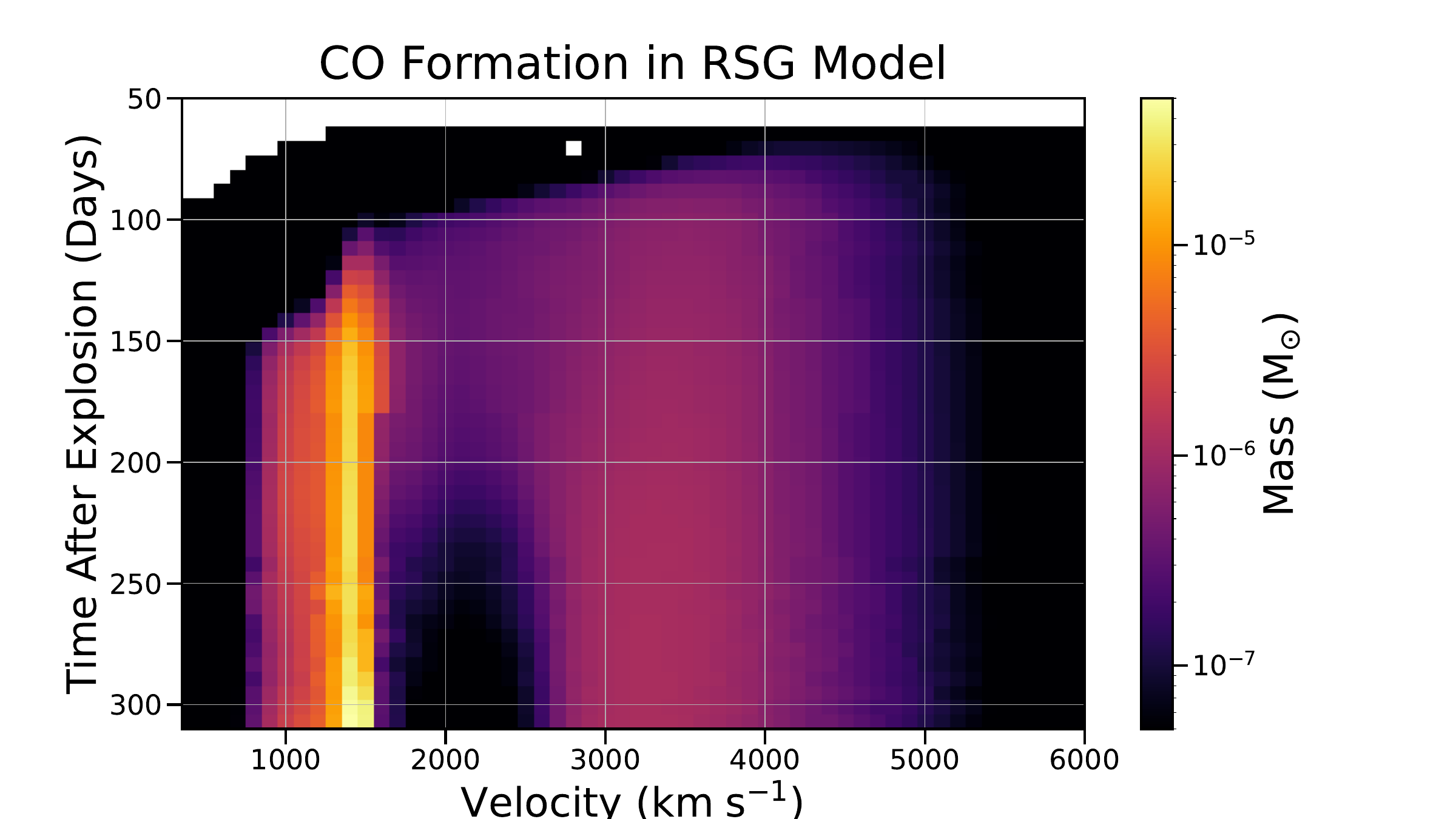}}
  \caption{Colormap of CO formation in our RSG model.  Each pixel is colored according to the total mass of CO present within a range of 100 \kms .  Most of the CO is concentrated in the dense O/Si-rich region at $\sim 1500$ \kms .  A secondary region of CO formation is apparent between $\sim 3000$ and $5000$ \kms , which is more extended and has lower density.  The two regions are separated by a $\sim 1000$ \kms\   wide region which is almost devoid of CO. The mass of CO exterior to 6000 \kms\   is negligible.}
  \label{fig:co_heatmap}
\end{figure}

In our RSG model, CO forms in two distinct locations, with a region of high temperature and low CO density separating them.  In both regions, cooling due to CO lines constitutes a significant portion ($>25\%$) of the total cooling.  Most (around $3/4$ of the total mass) of the CO forms in the oxygen/silicon-rich region at velocities <1500 \kms\, with the remaining CO spread over the broader oxygen/carbon-rich region extending from 2500 \kms\ to 4500 \kms .  The first region is not as cool but is denser, illustrating the conditions that lead to CO formation: high density, low temperature, or both. The CO band emission in the spectrum is dominated primarily by emission from the lower-velocity region, where most of the CO mass is formed. The separating region has intermediate density but lower carbon abundance and higher temperature, reducing the amount of CO that forms.  By nebular times, the two CO regions have cooled to a significantly lower temperature than the separating region.  Fig. \ref{fig:co_heatmap} shows the evolution of CO in our RSG model as a function of velocity and time. In our BSG model, all the CO is formed in the central regions, and the CO density quickly becomes negligible at velocities >1500 \kms .


\begin{figure}
  \centering
  \makebox[\linewidth][c]{\includegraphics[width=1.0\linewidth]{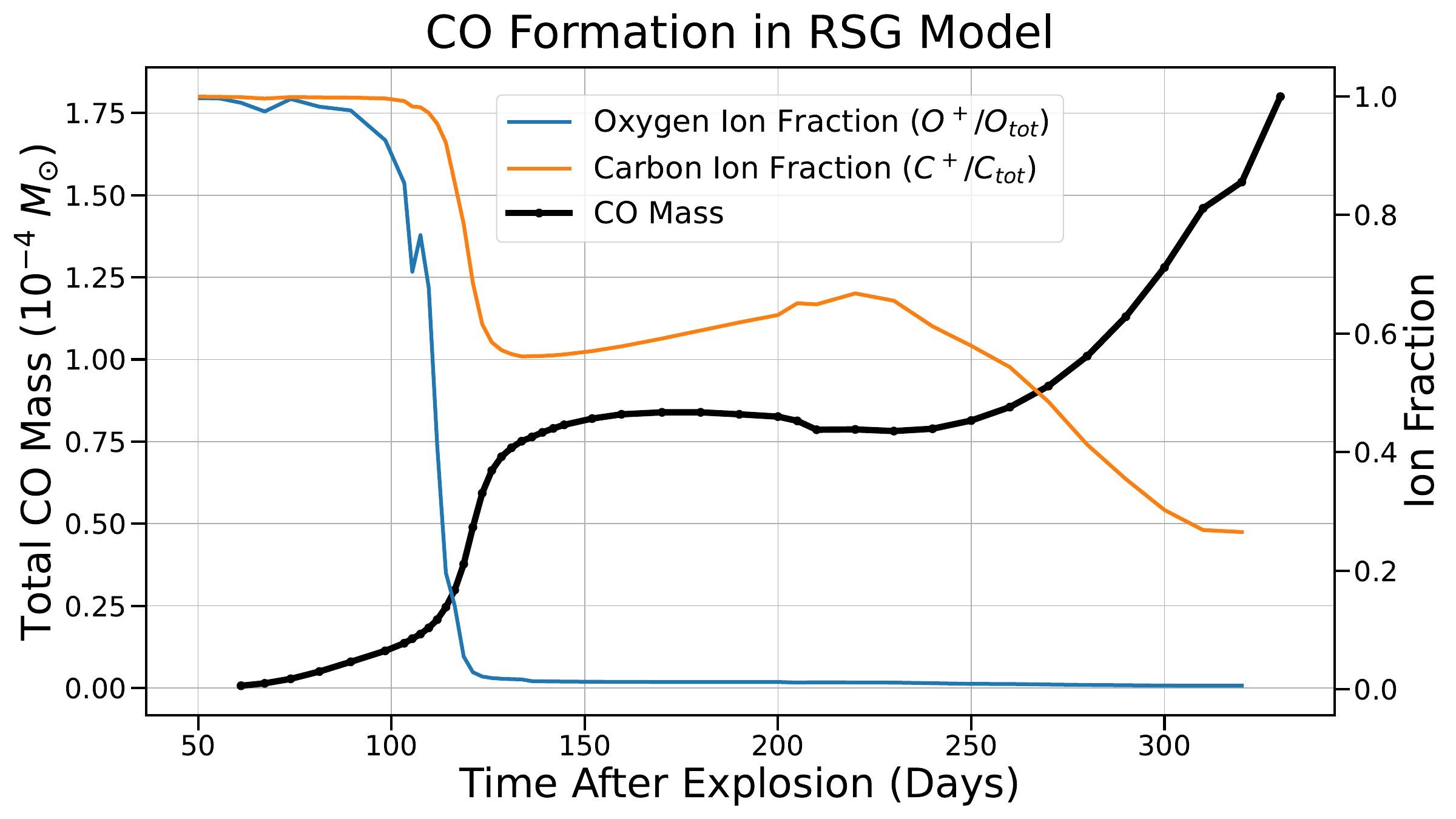}}
  \caption{Total mass of CO present in the RSG model (black) as a function of time since explosion. Also shown are ionization fractions for carbon (orange) and oxygen (blue) in the CO region. We calculate the ionization fractions as an average weighted by the final CO density: $\frac{\int (N^+/N_{\rm{tot}}) \times \rho_{CO} dV}{\int \rho_{CO} dV}$ (where $N_{\rm{tot}}$ is the total number density of oxygen [carbon] and $N^+$ the number density of singly-ionized oxygen [carbon]) to illustrate the importance of the ionization structure on CO formation. The amount of CO present is negligible before 60 days. Most of the CO is formed in a short window between $\sim 110$ and $\sim 130$ days post-explosion, as oxygen and carbon start to recombine.  The amount of CO reaches a maximum at $\sim 150$ days, before roughly plateauing as carbon remains ionized for some time. After 220 days, the mass of CO begins to rise again as neutral carbon becomes the dominant ionization stage.}
  \label{fig:co_mass_form}
\end{figure}

The amount of CO grows slowly between 50 and 100 days, after which its growth begins to accelerate (Fig. \ref{fig:co_mass_form}). The growth of CO then levels off at around 130 days post-explosion.  The earliest formation of CO (before 100 days) occurs primarily in the cooler formation region between 3000 and 4000 \kms, where the oxygen ionization fraction is lower than in the hotter region below 1500 \kms. Ionization of oxygen and carbon dictates when CO forms, as reactions involving neutral carbon and oxygen dominate CO formation. The initial burst of CO formation occurs as oxygen recombines, corresponding to the plateau drop-off of the SN light curve when the photosphere has moved through the hydrogen envelope.  The predominant formation pathway at this epoch is radiative association of neutral carbon and oxygen.  Carbon remains ionized for longer, causing the amount of CO to plateau between 130 and 220 days.  After 220 days, neutral carbon starts to become the dominant ionization stage, allowing for more CO formation.  At this stage, the growth of CO is also due to a pair of formation reactions which involve neutral oxygen as a reactant (reactions 13 and 27 in Table \ref{table:rxns}). At 300 days after explosion, the total mass of CO in our RSG model is around $10^{-4} M_{\odot}$.


\subsection{Spectral Features}

\begin{figure}
  \centering
  \makebox[\linewidth][c]{\includegraphics[width=1.0\linewidth]{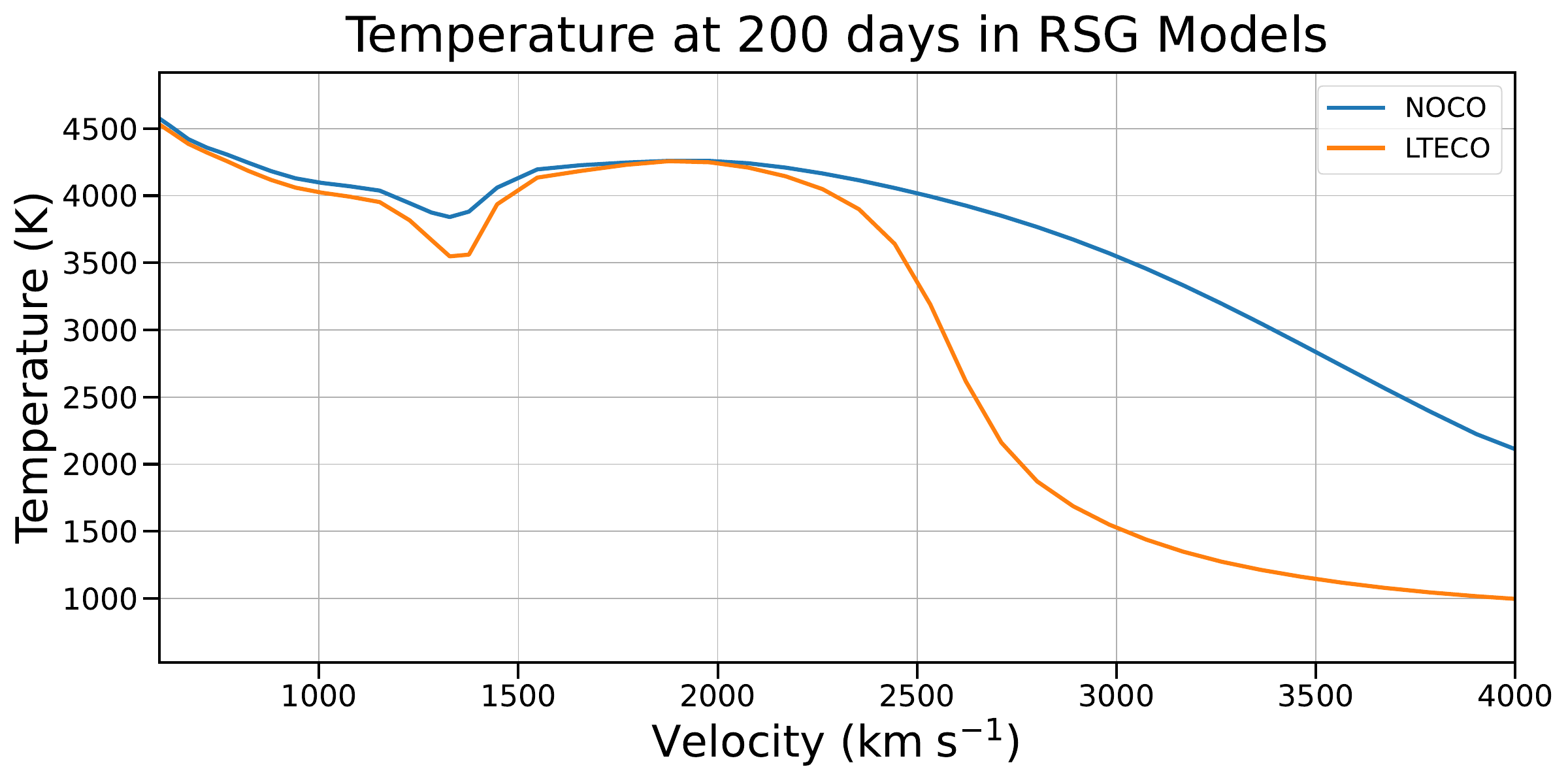}}
  \caption{Temperature structures in our RSG models at 200 days post-explosion. The two separate regions of CO formation are visible here as lower-temperature regions: a narrow one near 1400 \kms, and a broader region between 2500 and 4000 \kms. The presence of CO in these regions has accelerated the cooling of the ejecta significantly by 200 days.}
  \label{fig:temp_iip}
\end{figure}

The addition of line emission from CO adds a new efficient cooling channel, causing the nebular-phase equilibrium temperature in the CO region to be lower than without CO, by as much as a factor of two (Fig. \ref{fig:temp_iip}).  While the SN is still in the photospheric phase, CO emission does not significantly affect the spectrum, as very little CO forms before the end of the plateau phase.  As the SN evolves to the nebular phase, the impact of CO cooling becomes more pronounced: the lower temperature in the CO region reduces emissivity in the lines formed in that region, notably oxygen and calcium lines. In our mixed models, the reduced CO-region temperature depresses emission from all elements, but in more realistic unmixed models, it would be expected to impact only lines formed in the oxygen/carbon-rich region.


\begin{figure*}
  \centering
  \makebox[\linewidth][c]{\includegraphics[width=1.1\linewidth]{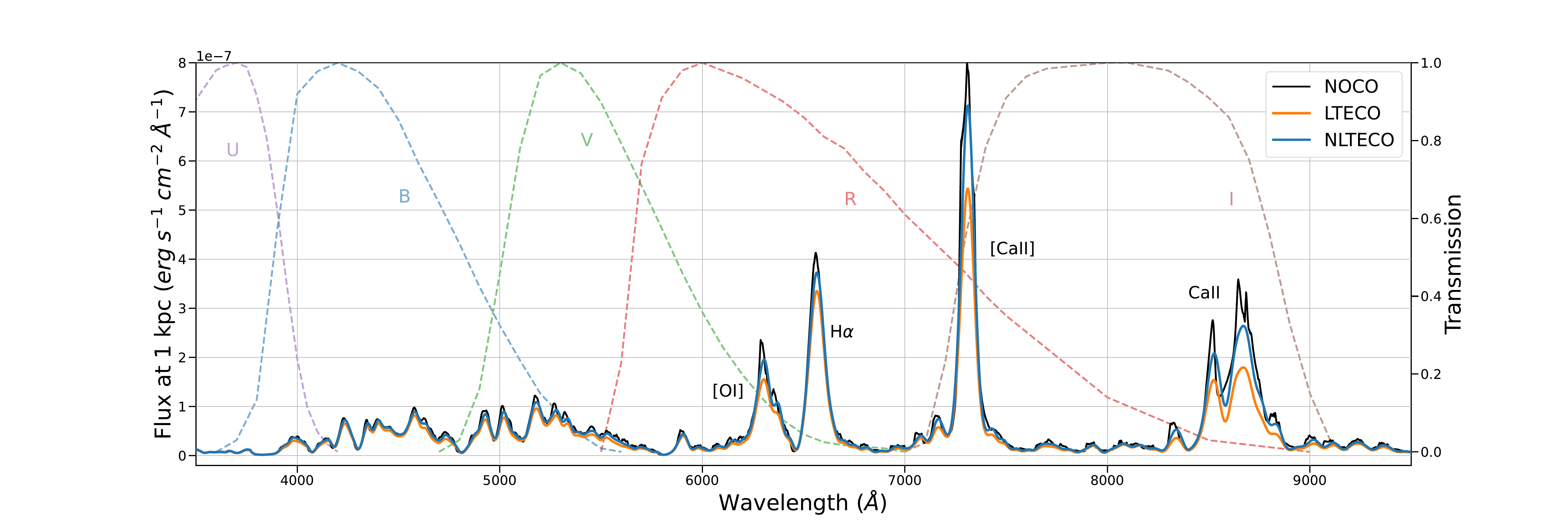}}
  \caption{Optical spectra computed from BSG models at 330 days without CO (NOCO), with CO level populations in LTE (LTECO), and with CO level populations allowed to depart from LTE (NLTECO). The strongest emission lines are identified. The CO models are cooler in the inner region ($<3500$ \kms), reducing the emissivity of lines formed there, with the most significant effect on \ion{Ca}{II} lines in the optical. Also shown are transmission curves for Johnson-Cousins optical filters \citep{bessell_standard_2005}.} 
  \label{fig:spec_filters_opt}
\end{figure*}

\begin{figure*}
  \centering
    \makebox[\linewidth][c]{\includegraphics[width=1.1\linewidth]{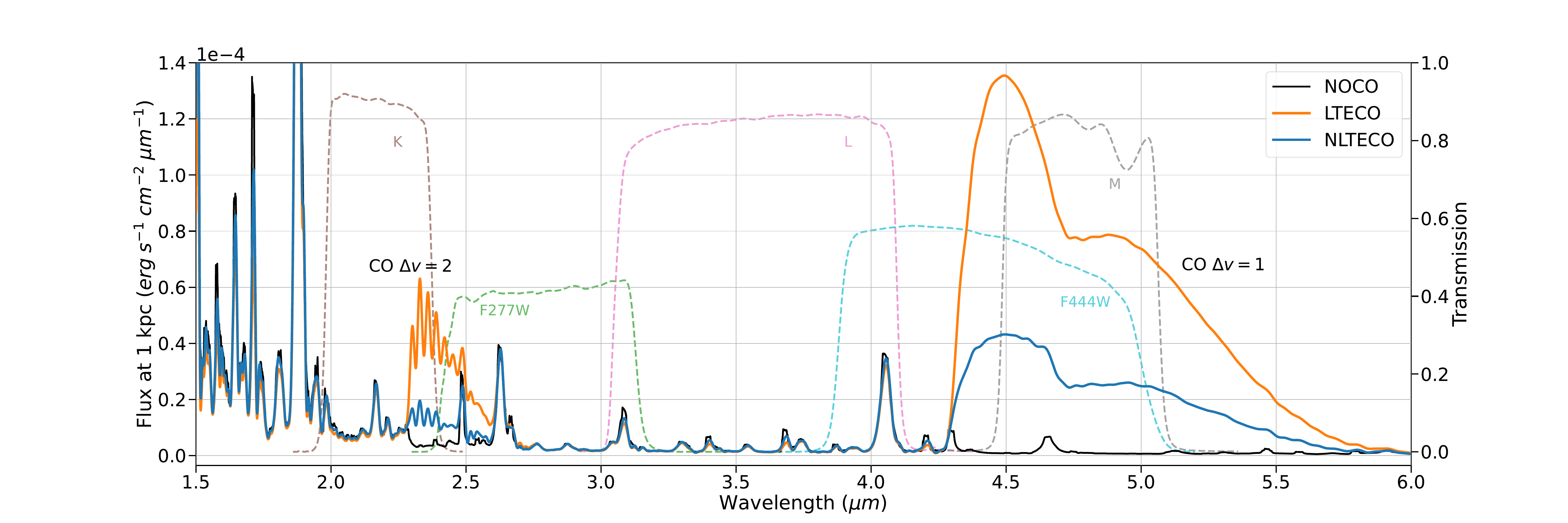}}
    \caption{Infrared spectra for the same BSG models as Fig. \ref{fig:spec_filters_opt}. Emission from CO primarily emerges in the first overtone ($\sim 2.5 \mu m$, $\Delta v = 2$) and fundamental bands ($\sim 5 \mu m$, $\Delta v = 1$). Emission from CO is reduced when level populations are allowed to depart from LTE, reducing the net cooling effect from CO. Also shown are some infrared filter transmission functions: J, H, and K bands used by the European Southern Observatory's Very Large Telescope (VLT) \citep{wamsteker_standard_1981}, and the James Webb Space Telescope's (JWST) NIRCam F277W and F444W bands \citep{rieke_overview_2005}, which cover the main regions of CO emission. The plotted curves for F277W and F444W include all mirror and detector effects, while the J, H, and K band curves account for only the filter transmission.}
    \label{fig:spec_filters_ir}
\end{figure*}

\begin{figure}
  \centering
  \makebox[\linewidth][c]{\includegraphics[width=1.1\linewidth]{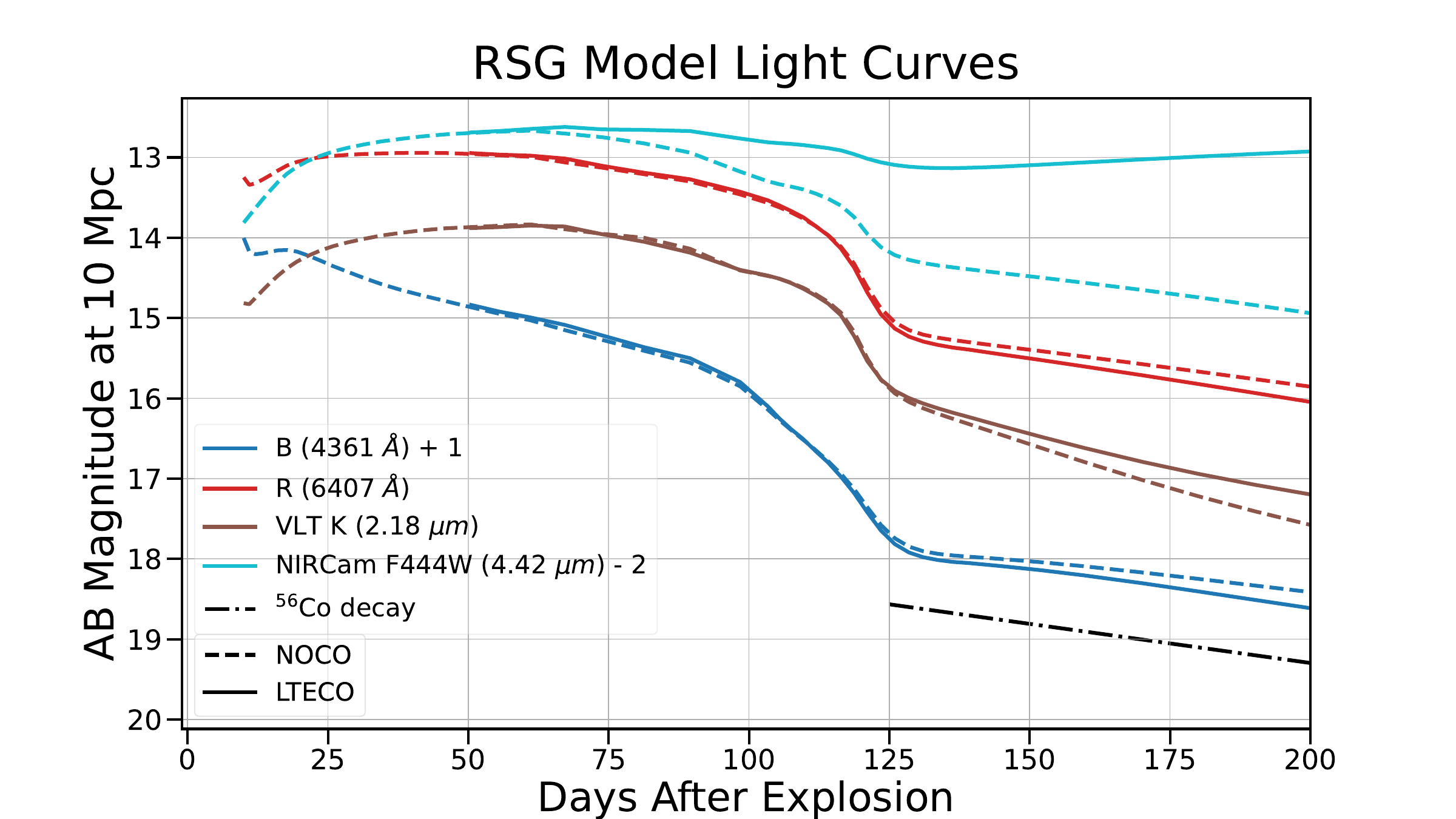}}
  \caption{Light curves derived from our RSG model with and without CO formation included. Effective wavelengths for each filter are indicated in the legend. Magnitudes are given as AB magnitudes observed at a nominal distance of 10 Mpc. In optical and NIR filters, the CO model decays more rapidly at late times as CO cooling begins to dominate.  In Mid-IR filters which overlap the CO emission bands (like F444W), the CO model is much brighter, with its brightness still increasing at 200 days.}
  \label{fig:light_curves}
\end{figure}

Optical spectra calculated from our BSG models at the nebular phase with and without CO are shown in Fig. \ref{fig:spec_filters_opt}. The impact of CO can be seen in all the lines, but is most pronounced in the strongest lines, including H$\alpha$ and CaII:  the effect on these features is highly dependent on our imposed mixing.  The vast majority of CO emission emerges in the first overtone and fundamental bands, with most of the emission in the fundamental band at $\sim 4.5 - 5.5 \mu m$ (Fig. \ref{fig:spec_filters_ir}).  Even with less than 1 per cent of carbon atoms in the form of CO, emission from the CO fundamental band accounts for as much as 20 per cent of the total luminosity of the CO model. The addition of cooling from CO lines means the ejecta cool faster than without CO, resulting in a faster-declining light curve at optical wavelengths (Fig. \ref{fig:light_curves}).  The radiation that would have been emitted in the optical is instead emitted in CO rovibrational bands in the infrared.


Broadband filters which overlap the CO emission in the infrared (Fig. \ref{fig:spec_filters_ir}) show significantly more flux at late times; brightness in the F444W NIRCam filter actually rises after 120 days as CO fundamental band emission becomes more important. The K band and some JWST bands overlap with CO emission bands in the infrared, and should be sensitive to CO emission in SNe. Mid-infrared photometric observations using Spitzer and WISE have shown evidence for CO emission in multiple SNe \citep{ergon_optical_2014,gandhi_sn_2013}, and JWST's greater sensitivity should improve constraints on CO emission. Most existing observational constraints on CO emission come from the first overtone, as in, e.g., SN2017eaw \citep{rho_near-infrared_2018}. JWST, however, can observe both the first overtone and fundamental bands of CO emission with high resolution; its upcoming observational programs could substantially improve our understanding of CO emission in SNe. The relative strength of the first overtone and fundamental bands depends on the optical depth. Constraining both bands simultaneously can improve the accuracy of observational estimates of CO mass (see \S \ref{sec:87a_obs}).

\begin{figure}
  \centering
  \makebox[\linewidth][c]{\includegraphics[width=1.1\linewidth]{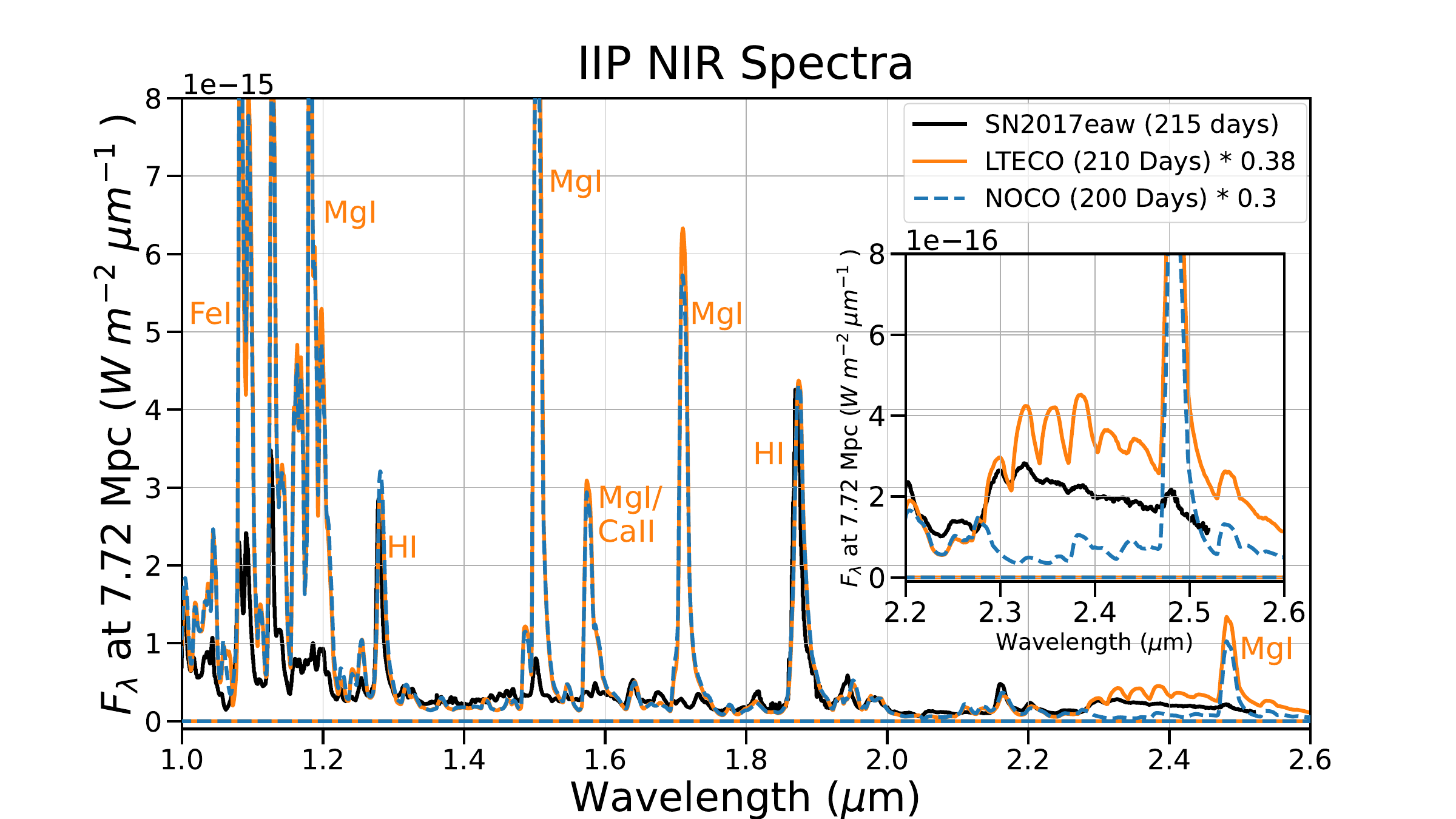}}
  \caption{NIR RSG model spectrum at 210 days with a NIR spectrum of Type IIP SN2017eaw taken at 215 days after explosion. The SN2017eaw spectrum is taken from \citet{rho_near-infrared_2018}.  The inset plot shows the CO first overtone emission band.  The model overestimates the CO flux, especially in the higher-excitation bandheads, despite containing less CO mass than was estimated for 2017eaw.  Some discrepancy is expected from our assumption of LTE level populations in CO. The model also significantly overestimates the strength of several \ion{Mg}{I} lines in this region, suggesting that the model's magnesium abundance may be much higher than SN2017eaw. The strength of the \ion{Mg}{I} lines is slightly increased when CO is included, as a result of lower ionization in the CO-dominated regions.}
  \label{fig:iip_nir_spec}
\end{figure}

Fig. \ref{fig:iip_nir_spec} shows our RSG model spectrum at 210 days compared to a similar-epoch near-infrared spectrum of SN2017eaw (from \citet{rho_near-infrared_2018}). Note that the CO bands in the emergent spectrum are dominated by emission from the lower-velocity formation region in our RSG model. \citet{rho_near-infrared_2018} estimated a CO mass of $\sim \; 2 \, \times \, 10^{-4} \, M_{\sun}$ from the first overtone emission at this epoch, roughly three times the CO mass present in our RSG model at 210 days. Despite this discrepancy, our model produces more emission in the first overtone than in the observations.  The relative bandhead strength in our model also differs from the observations, with the model overestimating the relative strength of the longer-wavelength bandheads (which come from higher-energy vibrational states). Our models (with and without CO) significantly overestimate the strength of several \ion{Mg}{I} lines present in this wavelength region, which could be due to errors in ionization or magnesium abundance. Magnesium lines in the infrared are sensitive to a number of progenitor properties, any of which could contribute to the discrepancy between our model and SN2017eaw. See \citet{dessart_radiative-transfer_2020} for a discussion of how properties including the progenitor hydrogen envelope mass, the extent of chemical mixing, or the progenitor ZAMS mass affect the ionization in the ejecta and the appearance of \ion{Mg}{I} lines.


\subsection{Clumping and Dust}

Clumping in SN ejecta consists of small regions of enhanced density (clumps) separated by much lower-density material in the inter-clump region.  The total amount of CO formed in the ejecta depends strongly on density--a clumpier ejecta profile would likely produce more CO and cool faster than uniform ejecta. \citet{dessart_impact_2018} investigated the effects of clumping in a similar \textsc{cmfgen} model of a Type II-peculiar SN, finding that the inclusion of clumping resulted in a faster-evolving light curve, redder colors, and an earlier transition to the nebular phase. The inclusion of both CO cooling and strongly clumped ejecta would likely allow for earlier formation of CO and exacerbate both the effects of clumping described in \citet{dessart_impact_2018} and the effects of CO cooling inferred in this work. In fact, results from multidimensional simulations suggest the ejecta should be most strongly clumped (i.e., higher densities within clumps) in the inner regions of the ejecta where CO is most important as a result of local expansion driven by ``nickel bubbles'' \citep{mueller_instability_1991,wongwathanarat_three-dimensional_2015}. 


The presence of CO has implications for dust formation by cooling the ejecta and potentially affecting the abundance of carbon atoms available to form dust.  The inclusion of CO speeds the rate of cooling in the densest ejecta, allowing for the formation of dust earlier.  Our model also predicts CO abundances significantly smaller than the total carbon or oxygen abundance (<1 per cent), leaving both species available to form larger molecular chains, i.e., dust.


\section{CO Cooling behavior}\label{sec:cooling}

\subsection{CO cooling function}

\begin{figure}
  \centering
  \makebox[\linewidth][c]{\includegraphics[width=1.1\linewidth]{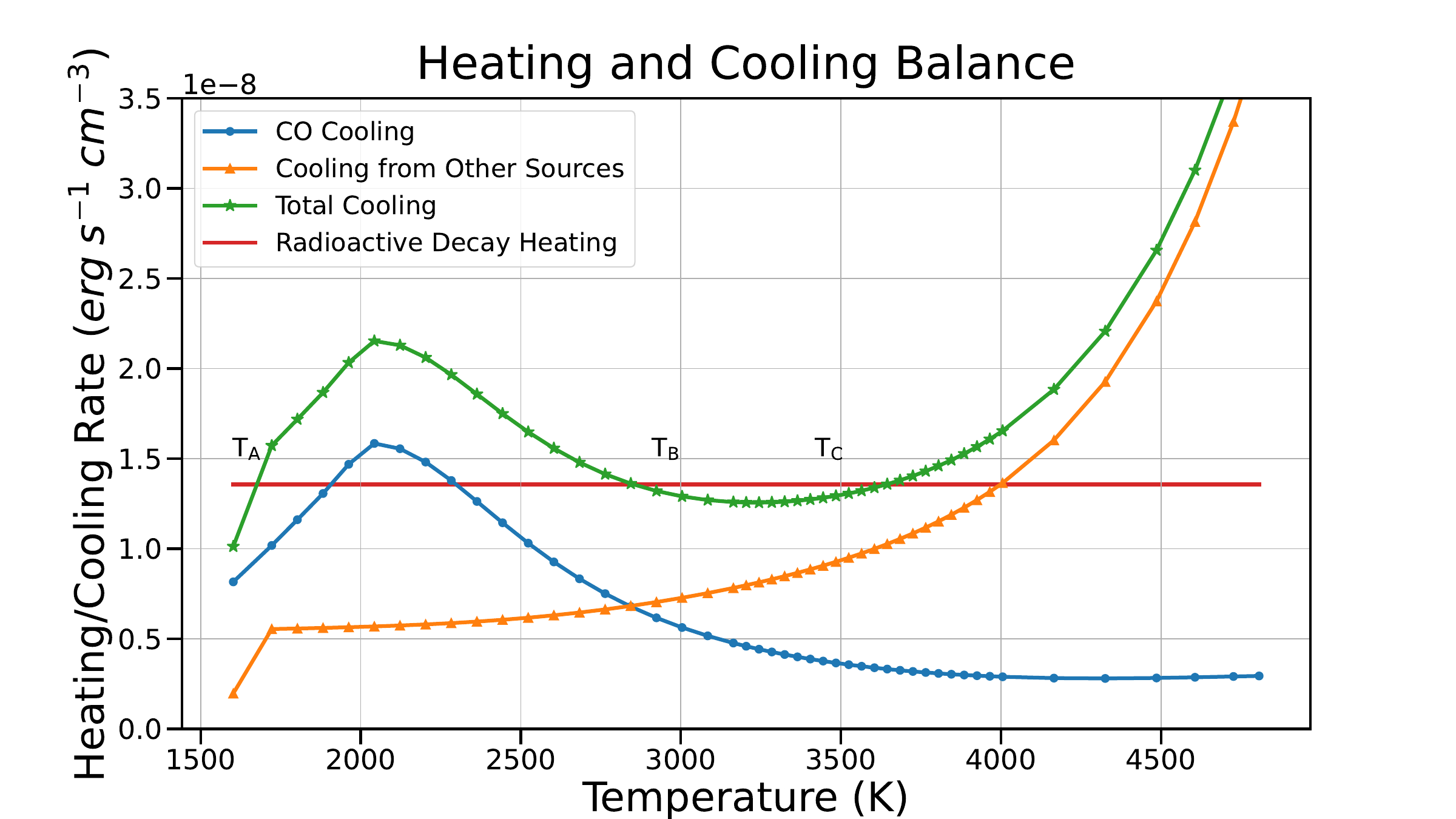}}
  \caption{Heating and cooling balance at a single region in our BSG model. The equilibrium temperature is where total cooling (green) equals total heating (red, from radioactive decay). The addition of CO cooling causes the cooling curve to rise at temperatures near 2000 K. The CO cooling drops off again below 2000K, with the result that the total cooling curve intersects the heating curve at up to 3 distinct temperatures, labeled $\mathrm{T}_\mathrm{A}$, $\mathrm{T}_\mathrm{B}$, and $\mathrm{T}_\mathrm{C}$.}
  \label{fig:valley}
\end{figure}

The CO cooling function has a non-monotonic dependence on temperature, due to the interaction of processes that depend on temperature in different ways.  CO molecules form preferentially at low temperatures, with the equilibrium CO abundance increasing by a factor of $\sim 30$ when the temperature decreases from 4000 to 2000 K.  However, CO emissivity per molecule is an increasing function of temperature, roughly proportional to $e^{-h \nu /k_B T}$.  The total CO cooling function, the product of these two processes, has a noticeable peak and decreases at higher and lower temperatures (Fig. \ref{fig:valley}).  The exact location of the peak also depends on the local density (density, temperature, and composition being the primary drivers of molecule formation), with the peak shifting to higher temperatures as density increases. We compute the total CO cooling rate by converging a model at fixed temperature, then summing the net radiative bracket times the photon energy (i.e., the net emission in the line) over all CO lines:

\begin{equation}
\int_{CO \, lines} N_u h \nu Z_{line} = \int_{CO lines} N_u h \nu \left( 1 - \frac{\bar{J}_{line}}{S_{line}} \right)
\end{equation}\label{eqn:net_bracket}

\noindent where $N_u$ is the upper state population, $Z_{line}$ the \textit{net radiative bracket}, $\bar{J}_{line}$ the frequency-averaged mean intensity, and $S_{line}$ the line source function. This represents the total cooling by CO line emission, taking into account non-local processes such as line overlap and radiative transfer effects.


The non-monotonic CO cooling function causes difficulties for finding the temperature solution in our models.  Some of the complications that can arise due to the shape of the CO cooling curve can be seen in Fig. \ref{fig:valley}.

Without CO cooling included, the total cooling curve monotonically increases with temperature.  When CO is included, it dominates the cooling behavior at temperatures around 2000-3000 K, forming a local maximum near 2000 K and a local minimum near 3200 K.  The new total cooling curve now crosses the heating curve at up to three points, each of which is a potentially valid temperature solution, labeled $\rm{T}_{\rm{A}}$, $\rm{T}_{\rm{B}}$, and $\rm{T}_{\rm{C}}$ in Fig. \ref{fig:valley}.  We can reject solution B on the grounds of stability, but solutions A and C are both stable.  Solution C lies closest to the temperature solution if CO cooling is not included, and could be considered the most physically realistic solution, as it is close to \citet{spyromilio_carbon_1988}'s estimate of the CO excitation temperature ($\sim$3000 K) in SN 1987A.


Multiple temperature solutions (hereafter scenario 1) can arise in models that include molecular cooling (Fig. \ref{fig:valley}).  Solutions A and C in Fig. \ref{fig:valley} are both physically valid; recall that the radiative-transfer problem comes with no guarantee of uniqueness. We expect that SN ejecta will evolve from high temperature to low temperature, so the initial temperature should be higher than $\rm{T}_{\rm{C}}$ and the gas should cool to that temperature without issue. However, the CO cooling phenomenon can cause more serious problems for our Newton-Raphson solution technique if the quantity of CO is large.

At some regions in our models, the density of CO is high enough that the local minimum in cooling at $\sim$3000 K is still greater than the heating rate, leaving solution A as the only temperature solution (scenario 2).  Depending on our initial estimate of temperature, our iterative Newton-Raphson solver can become trapped in this local minimum and fail to converge to a temperature solution without manual intervention.  Our initial temperature estimates, which typically come from similar models that do not include molecular cooling, lie at temperatures above $\mathrm{T}_\mathrm{C}$, and early iterations tend to fall into the local minimum.  Automating a correction for this effect, which would attempt to locate solution A, is non-trivial, as the location of the CO cooling peak depends on the local density, oxygen and carbon abundances, and the local radiation field. On the other hand, it may be the case that the density of CO is low enough that the local maximum near 2000 K lies below the heating curve (scenario 3).  In this scenario, only solution C is valid, and the local minimum is not a concern.  In this case, the CO cooling effect is small, and the final temperature solution is similar to that without CO.


\subsection{CO abundance and the cooling function}

The separation among the scenarios discussed above is not particularly large.  Because the total cooling curve does not vary strongly with temperature between 2000 and 4000 K, small changes in CO abundance can result in large changes in the inferred equilibrium temperature. If the CO abundance is increased by 50\%, the cooling curve in Fig. \ref{fig:valley} would shift to scenario 2, leaving only $\rm{T}_A$ as a temperature solution regardless of the initial conditions.  Conversely, a 50\% decrease in CO abundance moves it to scenario 3, leaving $\rm{T}_C$ as the only solution.  An equivalent change in the radioactive decay heating rate can also change the solution scenario.  Thus a factor of $\sim$2 change in the CO abundance, which is within the range of uncertainty in molecular reaction rates, separates the regime in which CO cooling is unimportant and the equilibrium temperature is $\sim$4000 K, to a regime in which CO cooling dominates and the equilibrium temperature is closer to 1500 K.


The equilibrium CO abundance depends sensitively on the rates of the most significant reactions that create and destroy it, notably radiative association (reaction 37 in Table \ref{table:rxns}, formation) and charge exchange with oxygen (reaction 9 in Table \ref{table:rxns}, destruction).  If one or both reaction rates are off by $\sim$50\%, it can alter not only the final temperature solution but the number of valid temperature solutions.  This is a significant concern, considering that most reaction rates available through UMIST \citep{mcelroy_umist_2013} are only accurate to within 25-50 per cent. Recent experimental results by \citet{meng_formation_2022} suggest the rate of radiative association may be as much as an order of magnitude higher than the UMIST rate we use in this work. Since this reaction typically dominates CO formation, an increased rate would significantly increase the amount of CO and associated cooling in our models. Using the radiative association reaction rate calculated by \citet{meng_formation_2022} increases the CO abundance by a factor of 10 and reduces the equilibrium temperature in the CO-dominated region of our BSG model by $\geq$500 K compared to the UMIST rate \citep{mcelroy_umist_2013}. Obtaining highly accurate reaction rates is thus of paramount importance, as deviations of a factor of 2 or less can change the temperature solution by thousands of Kelvin or introduce additional valid solutions.


Our assumptions of steady state and LTE level populations likely contribute to this issue. In a full time-dependent model, we expect the ejecta to evolve from high temperature to low temperature, and to settle into $\rm{T}_C$ without issue if the CO abundance is not too high. If CO level populations are allowed to deviate from LTE, then the net cooling from CO is reduced, potentially alleviating the problem of multiple solutions and leaving only the $\rm{T}_C$ solution. Some of our other assumptions, however, may be causing us to underestimate the importance of CO cooling. For example, unmixed models would be expected to produce higher CO densities in the oxygen/carbon shell, increasing the importance of CO cooling in that region and exacerbating this issue. Including strong clumping in the ejecta would also be expected to result in increased CO abundance and potentially lower-temperature solutions.  Alternatively, a change in our assumptions about how molecular reaction rates depend on the energy states of reactants (see \S \ref{sec:chem_rxns}) could have a sizeable impact on the CO abundance and subsequently the equilibrium temperature if a substantial portion of carbon or oxygen atoms are not in the ground state (as is likely the case in ionized gas).


While our BSG model suggests a total CO mass $\sim 1 \times 10^{-4} \; M_{\sun}$ at 330 days, observational estimates vary (\S \ref{sec:87a_obs}), mostly $\gtrsim 10^{-4} \; M_{\sun}$ \citep{spyromilio_carbon_1988,liu_oxygen_1995}, slightly above our model estimates (though not inconsistent, as the observational estimates come with large uncertainties and we utilize simple proof-of-concept models).

\section{Context}\label{sec:context}

This work is intended primarily to demonstrate the developments made to treat molecules in \textsc{cmfgen}, not reproduce observations, so detailed comparison to observations will be revisited in future works that make use of more realistic models. However, we make some comparisons both to observations of SN1987A and to other models of molecular formation to ensure that our results are not unreasonable.

\subsection{CO Observations in SN1987A}\label{sec:87a_obs}

\begin{figure*}
  \begin{center}
  \makebox[\linewidth][c]{\includegraphics[width=0.95\linewidth]{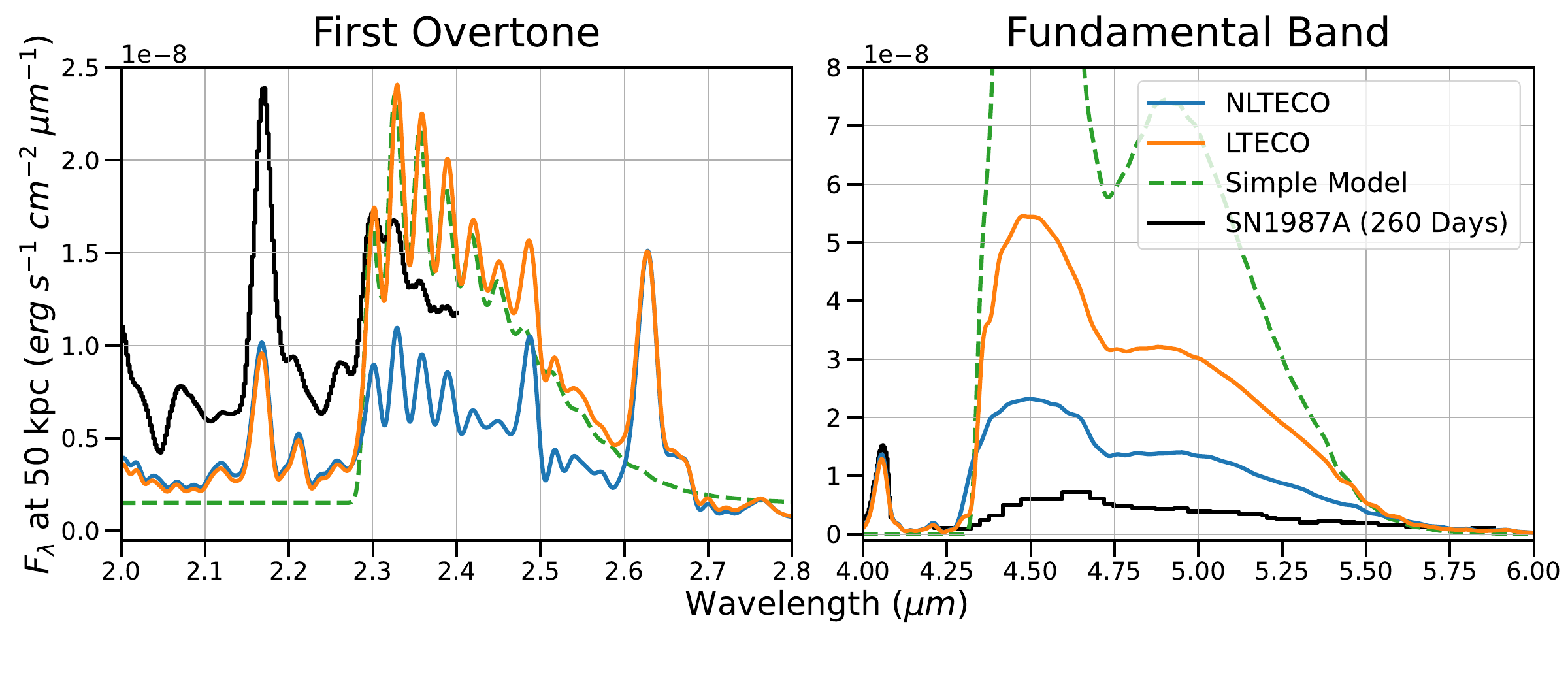}}
  \caption{Comparison of CO line appearance with and without the assumption of LTE.  The electron densities in our BSG model are below the critical density for the CO $v:0\rightarrow 1$ transition, resulting in excited-state populations well below the corresponding LTE populations.  This causes a factor $\sim 3$ reduction in flux in the fundamental and first overtone bands. In both cases, the first overtone lines remain optically thin ($\tau \sim 0.5$), while the fundamental band is optically thick ($\tau \sim 5$).  The ``simple model'' shown here assumes a spherical emission region of LTE, optically thin CO gas with total mass, max velocity, and excitation temperature chosen to match the \textsc{cmfgen} model properties, as used in \citet{spyromilio_carbon_1988}. It closely reproduces the first overtone band (in LTE), but overestimates even the LTE fundamental band emission by a factor of $\sim$4. Also shown is the spectrum of SN1987A at 260 days, scaled to match the change in bolometric magnitude between 260 and 330 days \citep{meikle_spectroscopy_1989,arnett_supernova_1989}. Our LTE model matches the flux in the first bandhead, but not the overall structure in the first overtone, and overestimates the fundamental band flux. Our NLTE model also overestimates the fundamental band flux despite underestimating the flux in the first overtone.}
  \label{fig:overtone_lte_and_nlte}
  \end{center}
\end{figure*}

The first detection of CO in a SN was in SN1987A, which exploded from a BSG progenitor in the Large Magellanic Cloud (LMC) in February 1987. Spectra of SN1987A taken as early as fall 1987 and spring 1988 showed evidence for CO fundamental band and first overtone emission, as well as SiO emission, with subsequent analysis deducing that the molecules must have formed within the ejecta \citep{arnett_supernova_1989}. \citet{spyromilio_carbon_1988} estimated the total mass of CO in the ejecta using a simple model which assumed a uniform, optically thin sphere of CO emitting in LTE.  We compare a similar model to the LTE first overtone emission in our BSG model at 330 days. The simple model, computed with total CO mass, maximum velocity, and excitation temperature chosen to match the properties of our BSG model, closely replicates the appearance of the first overtone band, suggesting that the first overtone region is optically thin in our model (see Fig. \ref{fig:overtone_lte_and_nlte}).  It does not accurately reproduce the fundamental band emission, however, which is optically thick in our BSG model.  When the assumption of LTE in our CO model is removed, the simple model fails to fit the appearance of either band---at this phase, the CO level populations are far from their LTE values. 


\citet{liu_carbon_1992} extended this simple model by also considering models without the assumptions of LTE populations or optical thinness. If CO populations deviate from LTE, the relative strength of shorter-wavelength bandheads, which correspond to radiative de-excitations from lower-energy vibrational states, should increase compared to longer-wavelength features (existing observations of SN1987A cover only the shortest wavelength bandheads in the first overtone band). In addition, higher optical depth in CO band regions can reduce the total emissivity at fixed CO mass, as photons are trapped until the transition is collisionally de-excited. This introduces a degeneracy when modeling only the first overtone band, as the feature could be produced by a large mass of CO with a high optical depth or a low mass with low optical depth. Using both the first overtone and the fundamental bands can potentially break this degeneracy, as the relative strength of the two should depend on the optical depth. In the optically thin limit, the total flux in the CO fundamental band should exceed the flux in the first overtone by about a factor of 20. 

In our models, regardless of the assumption of LTE, the fundamental band region is always optically thick ($\tau \sim 2-5$)\footnote{We calculate the optical depth in CO bands using direct integration in the observer's frame, accounting for the overlap of multiple rovibrational transitions.} and the fundamental/first overtone flux ratio is $< 20$. Observations of SN1987A at 260 days post-explosion from \citet{meikle_spectroscopy_1989} show a fundamental/first overtone flux ratio of $\sim 3$, suggesting that the optical depth in the fundamental band region is very high.


When using the same assumptions as in \citet{spyromilio_carbon_1988}, \citet{liu_carbon_1992} deduced significantly (5 to 10 times) higher CO masses at the same epochs, though the reason for the discrepancy is unclear. Our BSG model produces $\sim 10^{-4} \; M_{\sun}$ of CO at 330 days (regardless of the assumption of LTE), which is consistent with the \citet{liu_carbon_1992} estimate of the CO mass in SN1987A at 349 days when assuming LTE in an optically thin gas (and roughly consistent with the \citet{spyromilio_carbon_1988} estimate at the same epoch), but well below their estimate of $1.1 \times 10^{-3} \; M_{\sun}$ of CO at the same epoch when the assumption of LTE is removed.  The total mass of CO present in our BSG model is the same regardless of our assumption of LTE, but the strength of the CO bands changes dramatically (Fig. \ref{fig:overtone_lte_and_nlte}). Applying the method of \citet{spyromilio_carbon_1988} to the CO first overtone flux in our BSG LTECO model gives a mass estimate of $6.1 \times 10^{-5} M_{\odot}$, matching the mass of CO present in our model. Fitting our NLTE model flux with the same method suggests a CO mass of $2.0 \times 10^{-5} M_{\odot}$. Mass measurements made using this method are likely underestimated if the level populations are out of LTE.


When out of LTE, collisional processes determine CO level populations. Unfortunately, existing collisional data for CO rovibrational levels only cover vibrational transitions from $v=0$, requiring the use of theoretical scaling relations to cover all possible transitions \citep{ogloblina_electron_2020, chandra_collisional_2001}. Using this data, our NLTE models fail to reproduce the first overtone band appearance in either SN19897A or SN2017eaw, with our models consistently overestimating the strength of higher-order bandheads. This suggests that our treatment of NLTE level populations requires more accurate and comprehensive collisional data. Other features of the chosen model, such as progenitor mass, nature of chemical mixing, or composition may also contribute to the appearance of the CO bands.

Alternatively, the first overtone band appearance in SN1987A (Fig. \ref{fig:overtone_lte_and_nlte}, \citet{meikle_spectroscopy_1989}) can be fit by an LTE model with lower temperature, as in \citet{liu_carbon_1992,spyromilio_carbon_1988}. At lower excitation temperatures, higher vibrational levels will be less populated, reducing the strength of higher-order bandheads--i.e., de-excitations from higher vibrational states, which emit at longer wavelengths--and increasing the relative strength of the lowest-order bandheads (at shorter wavelengths). A simple model with an excitation temperature around 2000 K and $2 \times 10 ^{-4} M_{\odot}$ of CO fits the appearance of the first overtone at 260 days quite well \citep{liu_carbon_1992}. This temperature is also close to the very low-temperature solution we obtain in the CO-dominated region of our BSG LTECO model (\S \ref{sec:cooling}). However, such a model also predicts an even higher ratio ($\sim$50) of fundamental band to first overtone band flux, which is in tension with the observed ratio ($\sim$3) unless the fundamental band optical depth in 1987A is very high.
 
\subsection{Chemical Models}

Several studies have investigated CO formation in SN ejecta using chemical evolution models, including \citet{sluder_molecular_2018}, \citet{petuchowski_co_1989}, and \citet{liu_oxygen_1995}. Table \ref{tab:chem_mod_co_mass} shows a non-exhaustive list of CO masses computed using a variety of chemical evolution models for SN environments analogous to SN1987A.  Recent estimates of the CO mass tend to be somewhat higher than early estimates as a result of improved reaction networks and reaction rate data.  The earliest models in the table, \citet{petuchowski_co_1989} and \citet{lepp_molecules_1990}, assume fully-mixed ejecta and compute significantly lower CO masses due to charge exchange with He$^+$, which can quickly destroy CO molecules if present in the same region.  More recent works (including this one) neglect this reaction as a result of chemically stratified ejecta in which He and C/O reside in distinct shells, producing significantly more CO \citep{lepp_molecules_1990, gearhart_carbon_1999}.


We calculate somewhat lower total CO masses than contemporaneous works, which can be attributed partially to differences in chemical reaction networks, but is more likely a result of the strong mixing enforced in our BSG model, which results in lower C and O mass fractions in the inner regions where CO dominates as compared to an unmixed or macroscopically mixed model. Our NLTE BSG model predicts lower flux in the first overtone band than is observed in SN1987A, suggesting that our model could be underestimating the CO production. On the other hand, the same NLTE model still overestimates the strength of the CO fundamental band. Even $10^{-4} M_{\sun}$ of CO in our BSG model (total ejecta mass 13.2 $M_{\sun}$) is significant, with 20 per cent of the total flux emitted in CO bands in the LTE model, while the NLTE model sees the CO emission reduced by about a factor of 10 relative to the LTE model.  This mass of CO means that, in the C/O rich region of the ejecta, less than 1 per cent of C atoms are trapped in CO molecules, which allows for the later production of carbonaceous dust.


The most directly comparable studies to ours are those performed using \textsc{sumo}, described in \citet{liljegren_carbon_2020} and \citet{liljegren_molecular_2022}, which use a similar reaction network and include non-thermal and NLTE processes. We calculate lower CO masses than \citet{liljegren_carbon_2020}, most likely as a result of the model structure: they utilize a one-zone model consisting of only carbon and oxygen, while our BSG model features carbon and oxygen mixed throughout the ejecta with other elements. However, our individual reaction rates agree well with their results (with the exception of CO charge exchange with $\mathrm{O}^+$, which we find contributes significantly even after 200 days) and we find a similar impact from CO cooling, which reduces the temperature in the CO-dominated region (our models see a smaller cooling effect as a result of less CO). However \citet{liljegren_carbon_2020} do not report the same temperature solution issues described in \S \ref{sec:cooling} that result from CO cooling in our models. We also confirm the need for more accurate reaction rates, as the CO mass and equilibrium temperature depend strongly on the assumed rates: \citet{liljegren_carbon_2020} found that uncertainty in CO abundance scales roughly linearly with uncertainty in reaction rates. Since most rates in UMIST \citep{mcelroy_umist_2013} are accurate to about 50\%, our models can only constrain the CO abundance within 50\%, which covers all three temperature solution scenarios (\S \ref{sec:cooling}).

\begin{table*}
  \begin{center}
    \caption{CO masses computed at $\sim$300 days post-explosion for models of SN1987A and SN1987A-like models.  Differences among the models can be attributed to different assumptions about the ejecta composition, reaction networks, and reaction data.}
    \begin{tabular}{ c c c }
      \hline
      \multicolumn{3}{c} {CO mass in SN chemical evolution models} \\
      \hline
      \hline
      Study & CO Mass ($M_{\sun}$) & Epoch (Days After Explosion) \\
      \hline
      \citet{petuchowski_co_1989} & $2.5 \times 10^{-5}$ & 300 Days \\
      \citet{lepp_molecules_1990} & $4.1 \times 10^{-7} \; - \; 2.8 \times 10^{-5}$ & 349 Days \\
      \citet{liu_oxygen_1995} & $3.0 \times 10^{-3}$ & 300 Days \\
      \citet{cherchneff_molecules_2011} & $2 - 4 \times 10^{-3}$ & 300 Days \\
      \citet{sluder_molecular_2018} & $\sim 10^{-2}$ & 300 Days \\
      \citet{liljegren_carbon_2020} & $3 \times 10^{-3}$ & 300 Days \\
      \citet{ono_impact_2023} & $3 \times 10^{-2} \; - \; 2 \times 10^{-1}$ & 300 Days \\
      This Work & $\sim 10^{-4}$ & 330 Days \\
      \hline
    \end{tabular} \label{tab:chem_mod_co_mass}
  \end{center}
\end{table*}

\section{Conclusions}\label{sec:conclusion}

We have included a molecular reaction network in the radiative-transfer code \cmfgen\ to study the formation of CO and its impact on the gas and radiation properties in SN ejecta.  We account for the formation and destruction of CO by a number of chemical reactions and its impact on the ejecta via cooling by ro-vibrational emission. We investigate CO formation and cooling in a SN II-pec (BSG) model at 330 days post explosion and a SN IIP (RSG) model from 50 to 300 days post explosion, both of which feature microscopic mixing throughout the ejecta. We confirm that CO is an important coolant at late times in oxygen- and carbon-rich material.

In both models, CO forms in the densest and coolest regions of the ejecta, limited to low velocities (<2000 \kms ), in agreement with observations of CO. In our RSG model, CO becomes important (i.e., contributes a significant fraction of the total cooling) at $\sim$100 days after the explosion, and by $\sim$200 days after explosion, CO line cooling can constitute as much as 50 per cent of the total cooling in some regions.  The primary effect of CO cooling is to reduce the temperature by as much as two thousand Kelvin in CO-dominated regions. A reduced temperature weakens emission lines which are produced in the CO-dominated regions (due to the mixing in our models, we likely underestimate the impact on oxygen lines but overestimate the impact on lines from other species) and shifts that luminosity mostly to the CO fundamental band at $\sim 4.5 \mu m$. In this band, the ejecta remain optically thick well past 300 days. The CO first overtone band at $\sim 2.5 \mu m$ is nearly optically thin at 300 days--any estimates of CO mass made using these bands should take optical depth effects into account.

In our BSG model, CO dominates the total cooling in the inner regions. Both LTE and NLTE treatments of CO level populations significantly overestimate the ratio of fundamental band flux to first overtone band flux compared to observations of SN 1987A, potentially indicative of high optical depth in the CO fundamental band in SN1987A. Allowing CO level populations to deviate from LTE is necessary to reproduce the first overtone band structures in SN1987A (II-pec) and 2017eaw (IIP) \citep{meikle_spectroscopy_1989,rho_near-infrared_2018}. Our LTE models overestimate the strength of higher-order bandheads compared to observations of both SNe, suggesting the higher vibrational levels are underpopulated relative to LTE. Unfortunately, collisional excitation data for CO is limited, covering only a subset of possible transitions \citep{ogloblina_electron_2020}.

CO cooling presents numerical issues when determining the ejecta temperature structure. CO forms more efficiently in low-temperature environments, causing the total CO population to be a decreasing function of temperature. However, its emission per molecule is an increasing function of temperature. As a result, the CO cooling curve shows a strong peak between 2000 and 3000 K. In dense regions of the ejecta where CO cooling is important, the total cooling function becomes non-monotonic with a local minimum near 3000 - 4000 K.  The presence of a local minimum can trap numerical solvers or result in multiple temperature solutions.

The total amount of CO produced depends sensitively on the rates of a few formation and destruction mechanisms, predominantly radiative association (formation) and charge exchange with oxygen (destruction).  Literature values for the rates of these reactions come with large uncertainties (>25 per cent) and different sources report different rates: the rate \citet{dalgarno_radiative_1990} measured for radiative association differs by an order of magnitude from the rate measured by \citet{meng_formation_2022}.  Accurate rates are crucial for correctly estimating the mass of CO and the impact of CO cooling, and different rates can result in very different temperature solutions as a consequence of the non-monotonic CO cooling function (\S\ref{sec:cooling}).

Using the new capabilities of \textsc{cmfgen}, we will investigate the effects of CO cooling in SNe in more detail in future work. We plan to include CO in full time-sequence models of Type II-Plateau, Type II-peculiar, and Type Ibc SNe and explore macroscopic mixing and clumping. In addition, other potentially important molecules like silicon monoxide (SiO) and carbon monosulfide (CS), both of which have been observed in SN1987A \citep{meikle_spectroscopy_1993}, will be added to \cmfgen\ using the same approach detailed in this work.


\section*{Data Availability}


\cmfgen\ is available on Github or through the \cmfgen\ website at \url{https://sites.pitt.edu/~hillier/web/CMFGEN.htm}. The molecular version (including all molecular data utilized in this work) can be found on Github at \url{https://github.com/collinmcleod/cmfgen_mol}. Specific models utilized in this paper are available upon request.



\bibliographystyle{mnras}
\bibliography{mcleod_co_1}



\appendix

\section{Reaction Data}

    \begin{table*}
      \begin{center}
        \caption{Basic reaction data for all molecular reactions included in \textsc{cmfgen}. $\alpha$ has units of $cm^{-3} s^{-1}$, $\beta$ is dimensionless, and $T_{high}$, $T_{low}$, and $\gamma$ are in K. Rates are calculated using the specified constants for each reaction following Eqn. \ref{eqn:rate}. All data were obtained in 2021 from the UMIST Database for Astrochemistry \citep{mcelroy_umist_2013}.}
      \begin{tabular}{||c| c c c | c c c c c||}
        \hline
        ID & \multicolumn{3}{|c|}{Reaction} & $\alpha$ & $\beta$ & $\gamma$ & $T_{low}$ & $T_{high}$ \\
        \hline
        1 & $\ct + \ot^+$ & $\rightarrow$ & $\ct^+ + \ot$ & 4.10E-10 & 0 & 0 & 10 & 41000 \\
        2 & $\ct + \rm{O}^+$ & $\rightarrow$ & $\ct^+ + \rm{O}$ & 4.80E-10 & 0 & 0 & 10 & 41000 \\
        3 & $\rm{C} + \ct^+$ & $\rightarrow$ & $\rm{C}^+ + \ct$ & 1.10E-10 & 0 & 0 & 10 & 41000 \\
        4 & $\rm{C} + \co^+$ & $\rightarrow$ & $\rm{C}^+ + \co$ & 1.10E-10 & 0 & 0 & 10 & 41000 \\
        5 & $\rm{C} + \ot^+$ & $\rightarrow$ & $\rm{C}^+ + \ot$ & 5.20E-11 & 0 & 0 & 10 & 41000 \\
        6 & $\rm{C}^+ + \cto$ & $\rightarrow$ & $\rm{C} + \cto^+$ & 1.00E-9 & -0.5 & 0 & 10 & 41000 \\
        7 & $\cot^+ + \ot$ & $\rightarrow$ & $\cot + \ot^+$ & 5.30E-11 & 0 & 0 & 10 & 41000 \\
        8 & $\cot^+ + \rm{O}$ & $\rightarrow$ & $\cot + \rm{O}^+$ & 9.62E-11 & 0 & 0 & 10 & 41000 \\
        9 & $\co + \rm{O}^+$ & $\rightarrow$ & $\co^+ + \rm{O}$ & 4.90E-12 & 0.5 & 4.58E3 & 2000 & 10000 \\
        10 & $\co^+ + \ct$ & $\rightarrow$ & $\co + \ct^+$ & 8.40E-10 & 0 & 0 & 10 & 41000 \\
        11 & $\co^+ + \cot$ & $\rightarrow$ & $\co + \cot^+$ & 1.00E-9 & 0 & 0 & 10 & 41000 \\
        12 & $\co^+ + \ot$ & $\rightarrow$ & $\co + \ot^+$ & 1.20E-10 & 0 & 0 & 10 & 41000 \\
        13 & $\co^+ + \rm{O}$ & $\rightarrow$ & $\co + \rm{O}^+$ & 1.40E-10 & 0 & 0 & 10 & 41000 \\
        14 & $\rm{O}^+ + \ot$ & $\rightarrow$ & $\rm{O} + \ot^+$ & 1.90E-11 & 0 & 0 & 10 & 41000 \\
        15 & $\ct + \ot^+$ & $\rightarrow$ & $\co + \co^+$ & 4.10E-10 & 0 & 0 & 10 & 41000 \\
        16 & $\ct + \rm{O}^+$ & $\rightarrow$ & $\rm{C} + \co^+$ & 4.80E-10 & 0 & 0 & 10 & 41000 \\
        17 & $\ct^+ + \ot$ & $\rightarrow$ & $\co + \co^+$ & 8.00E-10 & 0 & 0 & 10 & 41000 \\
        18 & $\ct^+ + \rm{O}$ & $\rightarrow$ & $\rm{C} + \co^+$ & 3.10E-10 & 0 & 0 & 10 & 41000 \\
        19 & $\rm{C} + \ot^+$ & $\rightarrow$ & $\co^+ + \rm{O}$ & 5.20E-11 & 0 & 0 & 10 & 41000 \\
        20 & $\rm{C}^+ + \cot$ & $\rightarrow$ & $\co + \co^+$ & 1.10E-9 & 0 & 0 & 10 & 41000 \\
        21 & $\rm{C}^+ + \ot$ & $\rightarrow$ & $\co + \rm{O}^+$ & 4.54E-10 & 0 & 0 & 10 & 41000 \\
        22 & $\rm{C}^+ + \ot$ & $\rightarrow$ & $\co^+ + \rm{O}$ & 3.42E-10 & 0 & 0 & 10 & 41000 \\
        23 & $\cot + \rm{O}^+$ & $\rightarrow$ & $\co + \ot^+$ & 9.40E-10 & 0 & 0 & 10 & 41000 \\
        24 & $\cot^+ + \rm{O}$ & $\rightarrow$ & $\co + \ot^+$ & 1.64E-10 & 0 & 0 & 10 & 41000 \\
        25 & $\rm{C} + \cot$ & $\rightarrow$ & $\co + \co$ & 1.00E-15 & 0 & 0 & 10 & 41000 \\
        26 & $\ct + \ot$ & $\rightarrow$ & $\co + \co$ & 1.50E-11 & 0 & 4.3E3 & 298 & 1300 \\
        27 & $\ct + \rm{O}$ & $\rightarrow$ & $\rm{C} + \co$ & 2.00E-10 & -1.2E-1 & 0 & 10 & 8000 \\
        28 & $\cto + \rm{O}$ & $\rightarrow$ & $\co + \co$ & 8.59E-11 & 0 & 0 & 295 & 296 \\
        29 & $\rm{C} + \cto$ & $\rightarrow$ & $\co + \ct$ & 2.00E-10 & 0 & 0 & 10 & 300 \\
        30 & $\rm{C} + \co$ & $\rightarrow$ & $\ct + \rm{O}$ & 2.94E-11 & 5.0E-1 & 5.80250E4 & 1934 & 41000 \\
        31 & $\rm{C} + \ot$ & $\rightarrow$ & $\co + \rm{O}$ & 5.56E-11 & 4.10E-1 & -2.69E1 & 10 & 8000 \\
        32 & $\cot + \rm{O}$ & $\rightarrow$ & $\co + \ot$ & 2.46E-11 & 0 & 2.6567E4 & 300 & 6000 \\
        33 & $\co + \ot$ & $\rightarrow$ & $\cot + \rm{O}$ & 5.99E-12 & 0 & 2.4075E4 & 300 & 6000 \\
        34 & $\co + \rm{O}$ & $\rightarrow$ & $\rm{C} + \ot$ & 1.00E-16 & 0 & 0 & 1.0 & 1.1 \\
        35 & $\rm{C} + \rm{C}$ & $\rightarrow$ & $\ct + photon$ & 4.36E-18 & 3.5E-1 & 1.613E2 & 10 & 41000 \\
        36 & $\rm{C} + \rm{C}^+$ & $\rightarrow$ & $\ct^+ + photon$ & 4.01E-18 & 1.7E-1 & 1.015E2 & 10 & 41000 \\
        37 & $\rm{C} + \rm{O}$ & $\rightarrow$ & $\co + photon$ & 4.69E-19 & 1.52 & -5.05E1 & 10 & 14700 \\
        38 & $\rm{C} + \rm{O}^+$ & $\rightarrow$ & $\co^+ + photon$ & 5.00E-10 & -3.7 & 8.0E2 & 2000 & 10000 \\
        39 & $\rm{C}^+ + \rm{O}$ & $\rightarrow$ & $\co^+ + photon$ & 3.14E-18 & -1.5E-1 & 6.80E1 & 10 & 13900 \\
        40 & $\rm{O} + \rm{O}$ & $\rightarrow$ & $\ot + photon$ & 4.90E-20 & 1.58 & 0 & 10 & 300 \\
        41 & $\co^+ + e^-$ & $\rightarrow$ & $\rm{C} + \rm{O}$ & 2.00E-7 & -4.8E-1 & 0 & 10 & 41000 \\
        42 & $\cot^+ + e^-$ & $\rightarrow$ & $\co + \rm{O}$ & 3.80E-7 & -5.0E-1 & 0 & 10 & 300 \\
        43 & $\ct^+ + e^-$ & $\rightarrow$ & $\rm{C} + \rm{C}$ & 3.00E-7 & -5.0E-1 & 0 & 10 & 300 \\
        44 & $\cto^+ + e^-$ & $\rightarrow$ & $\co + \rm{C}$ & 3.00E-7 & -5.0E-1 & 0 & 10 & 300 \\
        45 & $\ot^+ + e^-$ & $\rightarrow$ & $\rm{O} + \rm{O}$ & 1.95E-7 & -7.0E-1 & 0 & 10 & 300 \\
        \hline
    \end{tabular} \label{table:rxns}
    \end{center}
    \end{table*} 
    %

\bsp	
\label{lastpage}
\end{document}